\documentclass[
   prd,
   amsfonts,
   amssymb,
   amsmath,
   superscriptaddress,
   twocolumn,
   eqsecnum,
   showpacs,
   preprintnumbers,
   byrevtex]{revtex4}
\usepackage{color}
\usepackage{graphicx}
\usepackage{subfigure}
\usepackage{time}
\newcommand{\beq}{\begin{equation}}
\newcommand{\bea}{\begin{eqnarray}}
\newcommand{\ee}{\end{equation}}
\newcommand{\eea}{\end{eqnarray}}

\newcommand{\er}{e_{r}}                    
\newcommand{\Er}{E_{r}}                    
\newcommand{\rd}{\mathrm{d}}               
\newcommand{\rD}{\mathrm{D}}               
\newcommand{\defby}{\equiv}                
\newcommand{\kperp}{k_{\perp}}             
\newcommand{\barkperp}{\bar{k}_{\perp}}    
\newcommand{\keta}{k_{\eta}}               
\newcommand{\peta}{\pi_{\eta}}               
\newcommand{\dpeta}{\dot{\pi}_{\eta}}        
\newcommand{\ddpeta}{\ddot{\pi}_{\eta}}      
\newcommand{\kphi}{k_{\phi}}               
\newcommand{\bx}{\mathbf{x}}               
\newcommand{\bk}{\mathbf{k}}               
\newcommand{\bP}{\mathbf{P}}               
\newcommand{\be}{\hat{\mathbf{e}}}         
\newcommand{\bsigma}{\boldsymbol{\sigma}}  
\newcommand{\Vb}{V}                        
\newcommand{\calE}{\mathcal{E}}                
\newcommand{\calPperp}{\mathcal{P}_{\perp}}    
\newcommand{\calPtheta}{\mathcal{P}_{\theta}}  
\newcommand{\calPeta}{\mathcal{P}_{\eta}}     
\newcommand{\calP}{\mathcal{P}}

\newcommand{\Diag}[1]{\mathrm{diag}( \, #1 \, )}   
\newcommand{\Set}[1]{( \, #1 \, )}                 
\newcommand{\Expect}[1]
   {\ensuremath{\langle \, #1 \,  \rangle}}
\newcommand{\Comm}[2]
   {\ensuremath{[ \, #1, #2 \, ]}}
\newcommand{\AntiComm}[2]
   {\ensuremath{\{ \, #1, #2 \, \}}}
\newcommand{\Tr}[1]{\mathrm{Tr} [ \, #1 \, ]}      
\newcommand{\Exp}[1]{\exp [ \, #1 \, ]}            
\newcommand{\Ln}[1]{\ln [ \, #1 \, ]}              
\newcommand{\Int}[1]{\int_{-\infty}^{+\infty}%
   \frac{ \mathrm{d} #1 }{ 2\pi } }                
\newcommand{\Ket}[1]{ | \, #1 \, \rangle }         
\newcommand{\BraKet}[2]%
   { \langle \, #1 \, | \, #2 \, \rangle }         
\begin{document}
%
%
%
%
\title[Pair production in (3+1) dimensions]
   {Backreaction and Particle Production in (3+1)-dimensional QED
   }

\author{Bogdan Mihaila}
\email{bmihaila@lanl.gov}
\affiliation{Materials Science and Technology Division,
   Los Alamos National Laboratory,
   Los Alamos, NM 87545, USA}

\author{Fred Cooper}
\email{cooper@santafe.edu}
\affiliation{National Science Foundation,
   4201 Wilson Blvd.,
   Arlington, VA 22230, USA}
\affiliation{Santa Fe Institute,
   Santa Fe, NM 87501, USA}
\affiliation{Center for Nonlinear Studies,
   Los Alamos National Laboratory,
   Los Alamos, New Mexico 87545, USA}

\author{John F. Dawson}
\email{john.dawson@unh.edu}
\affiliation{Department of Physics,
   University of New Hampshire,
   Durham, NH 03824, USA}



\begin{abstract}
We study the fermion pair production from a strong electric
field in boost-invariant coordinates in (3+1) dimensions and
exploit the cylindrical symmetry of the problem. This problem
has been used previously as a toy model for populating the
central-rapidity region of a heavy-ion collision (when we can
replace the electric by a chromoelectric field). We derive and
solve the renormalized equations for the dynamics of the mean
electric field and current of the produced particles, when the
field is taken to be a function only of the fluid proper time
$\tau = \sqrt{t^2-z^2}$. We determine the proper-time evolution
of the comoving energy density and pressure of the ensuing
plasma and the time evolution of suitable interpolating number
operators. We find that unlike in (1+1) dimensions, the energy
density $\varepsilon $  closely follows the longitudinal
pressure. The transverse momentum distribution of fermion pairs
at large momentum is quite different and larger than that
expected from the constant field result.
\end{abstract}
\maketitle
%
%
\section{Introduction}
\label{s:intro}

The ``Schwinger mechanism'' for pair production has been used
in various phenomenological models for particle production
following a high-energy heavy-ion collision.  One theoretical
picture of high-energy heavy-ion collisions begins with the
creation of a flux tube containing a strong color electric
field~\cite{r:Biro:1984kx}. The field energy is converted into
particles such as $q \bar{q}$ pairs and gluons by the Schwinger
mechanism~\cite{r:Sauter:1931kx,ref:HeisenbergEuler,r:Schwinger:1951fk}.
This mechanism has been  implemented in a phenomenological
fashion  in event generators for particle production such as
the Lund string model of hadronization~\cite{Lund} or the
Hijing model~\cite{Hijing}.  More recently another picture of
heavy-ion collisions, based on the color glass condensate model
of high density for quantum chromodynamics
(QCD)~\cite{colorglass1,colorglass2} has  been put forward.
This model  leads to the picture that a heavy-ion collision
produces an initial semi-classical chromoelectric field in the
longitudinal direction. Kharzeev \emph{et al.}~\cite{Kharzeev}
have shown that if one looks at a perturbative  parton cascade
model and studies inclusive production of gluons in a gluon
cascade, that this is equivalent to the production of a gluon
from a background classical chromoelectric field in the
longitudinal direction. This recent work  gives credence to the
idea that  as far as gluon production is concerned, one can
replace the dynamics of heavy-ion collisions by an initial
condition on a semiclassical chromoelectric field. In these
recent papers however, no attempt has been made to actually
study the time evolution of the  resulting plasma and the
backreaction of the production on the initial chromoelectric
field.   Some early studies had been done phenomenologically on
this type of problem using a kinetic theory model in which a
relativistic Boltzmann equation is coupled to a simple
Schwinger source
term~\cite{BC86,BC88,Kajantie,r:Gatoff:1987fk}, and a Wigner
function transport approach for an SU(2) version of QCD was
recently done by Skokov and Levai~\cite{su2}. A first principle
(quantum field theoretical)  calculation for pair production
and backreaction from strong fields  was done by one of us and
collaborators in the appropriate kinematics for heavy-ion
collisions in (1+1) dimensions in an abelian approximation
where one ignored the color degrees of freedom. The reason for
revisiting this problem now is two-fold. First, analytic
results for the transverse momentum distribution  functions for
particles produced by  constant electric and chromoelectric
fields have recently been
obtained~\cite{ref:NayakNei,r:Nayak:2005uq}. For the constant
chromoelectric field, the results for pair production are
different than for an electric field in that the transverse
distribution of jets depends not only on the energy density of
the field but also on the direction the field is pointing in
color space, i.e. the color hypercharge. Thus it is important
to know first, how the backreaction affects the transverse
momentum distribution function both for quantum electrodynamics
(QED) and QCD and secondly whether adding interactions in a
2-PI 1/N expansion will modify the one loop result.  Here we
will address the problem of finding the transverse distribution
function for the abelian case in (3+1) dimensions in a
realistic kinematic scenario. The QCD problem will be addressed
in a separate paper.

First let us review the history of both analytic approaches to
the constant electric field and chromoelectric field problem as
well as numerical studies of the backreaction problem. In his
1951 classic paper, Schwinger derived the following one-loop
nonperturbative formula
\begin{equation}\label{i.e:eq1}
   \frac{\rd W}{\rd^4 x}
   =
   \frac{e^2 E^2}{4\pi^3}
   \sum_{n=1}^{\infty}
   \frac{1}{n^2} \, e^{ -n\pi m^2 / |eE| }
\end{equation}
for the probability of fermion pair production per unit time
per unit volume from a constant electric field $E$ via vacuum
polarization~\cite{r:Schwinger:1951fk} by using a proper time
method.  The result of Schwinger was extended to QCD by
Claudson, Yildiz and Cox~\cite{r:Claudson:1980uq}.  However the
$p_T$ distribution of the $e^+$ (or $e^-$) production, $\rd W /
\rd^4x \, \rd^2 p_T$, could not be obtained using the proper
time method of Schwinger.  A WKB approximate method was used
for this purpose by Casher \emph{et
al.}~\cite{r:Casher:1979fk}, but an exact method to do this
problem (of determining the transverse distribution of pairs)
was not found until
recently~\cite{ref:NayakNei,r:Nayak:2005uq}.  For QED the WKB
analysis gave the correct answer which depended only on the
energy density of the electric field.  However, for QCD, the
WKB answer was  incorrect  for QCD in that it did not contain
the second Casimir invariant of $SU(3)$,
$C_2=[d_{abc}E^aE^bE^c]^2$, as shown in
Refs.~\onlinecite{ref:NayakNei}
and~\onlinecite{r:Nayak:2005uq}. In the case of fermions in QED
one finds for the transverse distribution of fermion pairs:
\begin{equation}\label{i.e:dsf}
   \frac{\rd W}{\rd^4x \, \rd^2 p_T}
   =
   -
   \frac{|eE|}{4\pi^3} \,
   \Ln{ 1 - e^{- \pi ( p_T^2 + m^2 ) / |eE| } } \>.
\end{equation}

The purpose of this paper is to consider the backreaction
problem in (3+1) dimensions in a situation which is related to
the kinematics of particle production by strong chromoelectric
fields, namely initial conditions where the center-of-mass
energy is so high that all distribution functions are boost
invariant in the longitudinal direction so that physical
quantities only depend on the longitudinal proper time which is
the same in the boost-invariant limit to the (1+1)-dimensional
fluid proper time.  Our goal is to see how the original result
of Casher \emph{et al.}~\cite{r:Casher:1979fk}, which has been
recently rigorously derived by Nayak for the transverse
distribution of fermion pairs~\cite{Nayak:2005uq}, is modified
by the expansion of the ensuing plasma and the backreaction on
the electric field.

The backreaction problem was first studied numerically in real
time for both scalar QED and QED by Cooper, Mottola and
collaborators in (1+1)
dimensions~\cite{r:CMprd40,r:KESCMprl91,r:Kluger:1992fk} and
then also in boost invariant coordinates relevant to heavy-ion
collisions in Ref.~\cite{r:Cooper:1993uq}. In
Ref.~\cite{r:Cooper:1993uq}, a strong abelian field was used as
a model for particle production in the central-rapidity
distribution. In that work, the boost invariance of the problem
was used to show that many features of the hydrodynamical model
were appearing even though there were no interactions kept that
would lead to  equilibration.  Also, in that paper, although
the theory was formulated in (3+1) dimensions, numerical
results were only presented for  (1+1) dimensions, so that the
transverse distribution of secondaries was not studied. To
remedy this particular deficiency of our previous work, here we
investigate the dynamics of the particle production as a
function of time in cylindrically-symmetric boost-invariant
coordinates in (3+1)-dimensional QED.

The present paper builds on our previous papers on fermion pair
production in (1+1) dimensions by strong electric fields with a
backreaction of the current on the
field~\cite{r:Cooper:1993uq,Mihaila:2008dp}. As in our previous
work, we employ quantum field theory methods in the large-N
approximation to find the particle production rate. The next
logical steps are to extend this result to QCD in this one-loop
approximation and then to do a self-consistent resummed 1/N
expansion to study the competition between the thermalization
of the plasma and the expansion. In this way we will build up
gradually the machinery to ask important questions about the
thermalization and expansion of the quark-gluon plasma in a
model based on the Schwinger mechanism.

The paper is organized as follows: In Sec.~\ref{s:theory} we
derive the equations needed for this calculation.  In
Sec.~\ref{s:energymomentum} we derive the components of the
energy-momentum tensor in this coordinate system and show that
it is conserved. In Sec.~\ref{s:quasi} we introduce the concept
of the quasiparticle phase space distribution function. This
quantity can be extracted from the field theory energy density
and then used to determine the distributions of pairs produced
in the center-of-mass frame. In Sec.~\ref{s:results} we discuss
our numerical approach and present results of our calculations.
We conclude in Sec.~\ref{s:conclusions}. In
App.~\ref{s:notation} we explain the notation we use throughout
this paper.  In App.~\ref{s:transhelicity} we derive the
transverse helicity eigenvectors we use in the main text of the
paper to expand the fermi field, whereas in
App.~\ref{s:adiabatic} we derive an adiabatic expansion of the
Dirac equation which are used throughout the paper to study the
large momentum behavior of integrands.

%
%
\section{Theory and notation}
\label{s:theory}

In cartesian coordinates $\xi^a = (\, t, x, y, z \,)$, and
using the matric $\eta_{ab} = \Diag{1,-1,-1,-1}$, the
lagrangian density for this problem is given by
\begin{multline}\label{lag}
   \mathcal{L}
   =
   \hat{\bar{\psi}}(\xi) \,
   \bigl \{ \,
      \gamma^a \,
      [ \,
         i \partial_a - e \, A_a(\xi) \,
      ]
      -
      M \,
   \bigr \} \, \hat{\psi}(\xi)
   \\
   -
   \tfrac{1}{4} \,
   F^{ab}(\xi) \, F_{ab}(\xi) \>,
\end{multline}
where $F^{ab}(\xi) = \partial^a A^b(\xi) - \partial^b
A^a(\xi)$.  Equations of motion are given by
\begin{gather}
   \bigl \{ \,
      \gamma^{a} \,
      [ \, i \partial_{a} - g \, A_{a}(\xi) \, ]
      -
      M \,
   \bigr \} \, \hat{\psi}(\xi)
   =
   0 \>,
   \label{e:eqI} \\
   \partial_{a} F^{ab}(\xi)
   =
   \langle \, \hat{j}^{b}(\xi) \, \rangle
   =
   g \,
   \Expect{
      \Comm{\hat{\bar{\psi}}(\xi)}{\gamma^{b} \, \hat{\psi}(\xi)}
          } / 2 \>.
   \label{e:eqII}
\end{gather}
In this semiclassical approximation, we quantize the Dirac
field while the electromagnetic field is treated classically.
This approximation can be made precise by considering $N$
flavors of quarks interacting with the electromagnetic field
and considering the limit where $N \rightarrow \infty$ after
appropriate re-scalings.  Systematic corrections are given by
the $1/N$ expansion as discussed in
Ref.~\onlinecite{r:CHKMPAprd94} and references therein.  We use
the standard representation of the $\gamma$-matrices
\begin{equation}\label{e:gammaSR}
   \gamma^0
   =
   \begin{pmatrix}
      1 & 0 \\
      0 & -1
   \end{pmatrix} \>,
   \qquad
   \gamma^i
   =
   \begin{pmatrix}
      0 & \sigma^i \\
      - \sigma^i & 0
   \end{pmatrix} \>,
\end{equation}
where $\sigma^i$ are the usual Pauli matrices, and $1$ is the unit
($2\times 2$) matrix.
%
%
\subsection{Dirac's equation in boost-invariant coordinates}
\label{ss:boostinvariant}

Boost-invariant coordinates $x^{\mu} = (\, \tau, \rho, \theta, \eta \,)$ are defined by
\begin{alignat}{2}
   t & = \tau \cosh \eta \>, & \qquad z & = \tau \sinh \eta \>,
   \label{e:xdefs} \\
   x & = \rho \cos\theta \>, & \qquad y & = \rho \sin\theta \>.
   \notag
\end{alignat}
We use Roman indices to indicate the cartesian frame and Greek
indices for the cylindrical-hyperbolic frame.  The connection
between the cartesian frame ($\rd \xi^a$), and the
boost-invariant frame  ($\rd x^{\mu}$) is described by a
vierbein matrix $\Vb^{a}{}_{\mu}(x)$, which for our case is
\begin{gather}
   \rd \xi^{a}
   =
   \Vb^{a}{}_{\mu}(x) \, \rd x^{\mu} \>,
   \qquad
   \partial_{\mu}
   =
   \Vb^{a}{}_{\mu}(x) \, \partial_{a} \>,
   \label{e:Vdef}
\end{gather}
with
\begin{gather}
   \Vb^{a}{}_{\mu}(x)
   \defby
   \frac{\partial \xi^{a}}{\partial x^{\mu}}
   =
   \begin{pmatrix}
      \cosh \eta & 0 & 0 & \tau \sinh \eta \\
      0 & \cos\theta & -\rho \sin\theta & 0 \\
      0 & \sin\theta & \rho \cos\theta & 0 \\
      \sinh \eta & 0 & 0 & \tau \cosh \eta
   \end{pmatrix} \>.
   \notag
\end{gather}
The inverse vierbein matrix, which we write as $\Vb^{\mu}{}_{a}(x)$, is given by
\begin{gather}
   \rd x^{\mu}
   =
   \Vb^{\mu}{}_{a}(x) \, \rd \xi^{a} \>,
   \quad
   \partial_{a}
   =
   \Vb^{\mu}{}_{a}(x) \, \partial_{\mu} \>,
   \label{e:Vinvdef}
\end{gather}
with
\begin{align}
   &
   \Vb^{\mu}{}_{a}(x)
   \defby
   \frac{\partial x^{\mu}}{\partial \xi^{a}}
   \notag \\ &
   =
   \begin{pmatrix}
      \cosh \eta & 0 & 0 & - \sinh \eta \\
      0 & \cos\theta & \sin\theta & 0 \\
      0 & -\sin\theta / \rho & \cos\theta / \rho & 0 \\
      - \sinh \eta / \tau & 0 & 0 & \cosh \eta / \tau
   \end{pmatrix} \>.
\end{align}
The metric $g_{\mu\nu}(x)$ in boost-invariant coordinates is
\begin{align}
   g_{\mu\nu}(x)
   &=
   \eta_{ab} \, \Vb^{a}{}_{\mu}(x) \, \Vb^{b}{}_{\nu}(x)
   \label{e:gi} \\
   &=
   \Diag{ 1, -1, -\rho^2, - \tau^2 } \>.
   \notag
\end{align}
We raise and lower Latin indices by the $\eta$-metric and Greek
indices by the $g$-metric.  So Dirac's equation in
boost-invariant coordinates can be written as
\begin{equation}\label{e:diracii}
   \bigl \{ \,
      \tilde{\gamma}^{\mu}(x) \,
      [ \,
         i \, \partial_{\mu} - e \, A_{\mu}(x) \,
      ]
      -
      M \,
   \bigr \} \,
   \hat{\psi}(x)
   =
   0 \>,
\end{equation}
where we have defined $\tilde{\gamma}^{\mu}(x) = \gamma^{a} \,
\Vb^{\mu}{}_{a}(x)$.  In this coordinate system, the
$\gamma$-matrices are given by
\begin{equation}\label{e:gammataueta}
\begin{split}
   \tilde{\gamma}^{\tau}(x)
   &=
      \cosh\eta \, \gamma^0
      -
      \sinh\eta \, \gamma^3 \,
   \>,
   \\
   \tilde{\gamma}^{\rho}(x)
   &=
      \cos\theta \, \gamma^1
      +
      \sin\theta \, \gamma^2 \,
   \>,
   \\
   \tilde{\gamma}^{\theta}(x)
   &=
   \bigl ( \,
      -
      \sin\theta \, \gamma^1
      +
      \cos\theta \, \gamma^2 \,
   \bigr ) / \rho \>,
   \\
   \tilde{\gamma}^{\eta}(x)
   &=
   \bigl ( \,
      -
      \sinh\eta \, \gamma^0
      +
      \cosh\eta \, \gamma^3 \,
   \bigr ) / \tau \>.
\end{split}
\end{equation}
The fermi field $\hat{\psi}(x)$ in boost-invariant coordinates
obeys the anticommutation relation
\begin{multline}\label{bi.e:psifieldanticomm}
   \AntiComm{
      \hat{\psi}_{\alpha}^{\phantom(}(\tau,\rho,\theta,\eta) }
            {
      \hat{\bar{\psi}}_{\alpha'}^{\dagger}(\tau,\rho',\theta',\eta') }
   \\
   =
   \tilde{\gamma}^{\tau}_{\alpha,\alpha'}(\eta) \,
   \frac{ \delta(\rho - \rho' ) }{ \sqrt{\rho \rho'} } \,
   \delta(\theta - \theta') \,
   \frac{ \delta( \eta - \eta' ) }{ \tau } \>.
\end{multline}

It is simpler, however, to solve Dirac's equation in a
Lorentz-transformed frame which diagonalizes the vierbein.  The
Lorentz transformation that does this is
\begin{equation}\label{e:VVandLambda}
\begin{split}
   \Lambda^{a}{}_{b}(\theta,\eta)
   &=
   \Vb^{a}{}_{\mu}(\tau,\rho,\theta,\eta) \,
   \bar{\Vb}^{\mu}{}_{b}(\tau,\rho)
   \\
   &=
   \begin{pmatrix}
      \cosh \eta & 0 & 0 & \sinh \eta \\
      0 & \cos\theta & -\sin\theta & 0 \\
      0 & \sin\theta & \cos\theta & 0 \\
      \sinh \eta & 0 & 0 & \cosh \eta
   \end{pmatrix} \>,
\end{split}
\end{equation}
where $\bar{\Vb}^{\mu}{}_{b}(\tau,\rho)$ are the diagonal vierbeins given by
\begin{align}\label{e:barVdefs}
   \bar{\Vb}^{a}{}_{\mu}(\tau,\rho)
   &\defby
   \Vb^{a}{}_{\mu}(\tau,\rho,0,0)
   =
   \Diag{ 1, 1, \rho, \tau }
   \\
   \bar{\Vb}^{\mu}{}_{a}(\tau,\rho)
   &\defby
   \Vb^{\mu}{}_{a}(\tau,\rho,0,0)
   =
   \Diag{ 1, 1, 1/\rho, 1/\tau } \>.
   \notag
\end{align}
We define $\gamma$-matrices in this frame with a bar
\begin{equation}\label{e:gammabardef}
   \bar{\gamma}^{\mu}(\tau,\rho)
   =
   \bar{\Vb}^{\mu}{}_{a}(\tau,\rho) \, \gamma^{a}
   =
   \tilde{\gamma}^{\mu}(\tau,\rho,0,0) \>.
\end{equation}
They are given explicitly by:
\begin{alignat}{2}
   \bar{\gamma}^{\tau}
   & =
   \gamma^{0} \>,
   & \qquad
   \bar{\gamma}^{\eta}(\tau)
   & =
   \gamma^{3} / \tau \>,
   \label{e:Vbardefs} \\
   \bar{\gamma}^{\rho}
   & =
   \gamma^{1} \>,
   & \qquad
   \bar{\gamma}^{\theta}(\rho)
   & =
   \gamma^{2} / \rho \>.
   \notag
\end{alignat}
Now let $S(\theta,\eta)$ be an operator which induces this
Lorentz transformation on the $\gamma^a$ matrices in the
orthogonal frame,
\begin{equation}\label{e:SgammaSL}
   S^{-1}(\theta,\eta) \, \gamma^{a} \, S(\theta,\eta)
   =
   \Lambda^{a}{}_{b}(\theta,\eta) \, \gamma^{b} \>.
\end{equation}
Then it is easy to show that $S(\theta,\eta) = S_{\rho}(\theta)
\, S_{\rho}(\eta)$ is given by a product of operators, where
\begin{subequations}\label{e:Srhotau}
\begin{align}
   S_{\rho}(\theta)
   &=
   \Exp{ \theta \, \gamma^1 \gamma^2 / 2 }
   \label{e:Srho} \\
   &=
   \cos( \theta/2 ) + \gamma^1 \gamma^2 \, \sin( \theta /2 ) \>,
   \notag \\
   S_{\tau}(\eta)
   &=
   \Exp{ \eta \, \gamma^0 \gamma^3 / 2 }
   \label{e:Stau} \\
   &=
   \cosh( \eta/2 ) + \gamma^0 \gamma^3 \, \sinh( \eta /2 ) \>.
   \notag
\end{align}
\end{subequations}
Furthermore, from Eq.~\eqref{e:VVandLambda}, we see that
$S(\theta,\eta)$ transforms the
$\tilde{\gamma}^{\mu}(\tau,\rho,\theta,\eta)$ matrices into the
$\bar{\gamma}^{\mu}(\tau,\rho)$,
\begin{equation}\label{e:SgammaSi}
   S^{-1}(\theta,\eta) \,
   \tilde{\gamma}^{\mu}(x) \,
   S(\theta,\eta)
   =
   \bar{\gamma}^{\mu}(\tau,\rho) \>.
\end{equation}
Now let us note that
\begin{multline}\label{e:SgammapS}
   S^{-1}(\theta,\eta) \,
   \tilde{\gamma}^{\mu}(x) \,
   \partial_{\mu} \,
   S(\theta,\eta)
   \\
   =
   \bar{\gamma}^{\mu}(\tau,\rho) \,
   \bigl [ \,
      \partial_{\mu}
      +
      \Pi_{\mu}(\theta,\eta) \,
   \bigr ] \>,
\end{multline}
where we have defined a connection $\Pi_{\mu}(\theta,\eta)$ by
\begin{equation}\label{e:Gammadef}
   \Pi_{\mu}(\theta,\eta)
   =
   S^{-1}(\theta,\eta) \,
   ( \, \partial_{\mu} S(\theta,\eta) \, ) \>.
\end{equation}
The only nonzero connections are when $\mu = \theta$ and
$\mu=\eta$. So using Eq.~\eqref{e:Srhotau}, we find
\begin{equation}\label{e:Gammarhoeta}
   \Pi_{\theta}
   =
   \gamma^1 \gamma^2 / 2 \>,
   \qquad\text{and}\qquad
   \Pi_{\eta}
   =
   \gamma^0 \gamma^3 / 2 \>,
\end{equation}
which are independent of $\theta$ or $\eta$.  The covariant
derivative $\nabla_{\mu}$ is given by
\begin{equation}\label{e:defcovder}
   \nabla_{\mu}
   \defby
   D_{\mu}
   +
   \Pi_{\mu}(x)
   \defby
   \partial_{\mu}
   +
   \Pi_{\mu}(x)
   +
   i e \, A_{\mu}(x) \>.
\end{equation}
Christoffel symbols for boost-invariant coordinates, which we will need later, are given by
\begin{equation}\label{e:christoffeldef}
   \Gamma_{\mu\nu}^{\lambda}(x)
   =
   \Vb^{\lambda}{}_{a}(x) \
   \partial_{\mu} \Vb^{a}{}_{\nu}(x) \>,
\end{equation}
from which we find the only non-vanishing elements to be
\begin{gather}
   \Gamma_{\eta\eta}^{\tau}
   =
   \tau \>,
   \qquad
   \Gamma_{\tau\eta}^{\eta}
   =
   \Gamma_{\eta\tau}^{\eta}
   =
   1 / \tau \>,
   \label{e:christoff} \\
   \Gamma_{\theta\theta}^{\rho}
   =
   - \rho \>,
   \qquad
   \Gamma_{\rho\theta}^{\theta}
   =
   \Gamma_{\theta\rho}^{\theta}
   =
   1 / \rho \>.
   \notag
\end{gather}

So Dirac's equation \eqref{e:diracii} can be transformed to the boost-invariant frame by defining
\begin{equation}\label{e:psitophi}
   \hat{\psi}(x)
   =
   S(\theta,\eta) \,
   \hat{\phi}(x) / \sqrt{\tau} \>,
\end{equation}
and multiplying the equation through by $S^{-1}(\theta,\eta)$, which gives the equation
\begin{equation}\label{e:diraciii}
   \bigl [ \,
      i \,
      \bar{\gamma}^{\mu}(\tau,\rho) \,
      \nabla_{\mu}
      -
      M \,
   \bigr ] \,
   \hat{\phi}(x) / \sqrt{\tau}
   =
   0 \>.
\end{equation}
For our case, we assume that the vector potential is in the
$\eta$-direction and depends only on $\tau$, so we choose
$A_{\mu}(x) = (\, 0,0,0, -A(\tau) \,)$, which defines $A(\tau)$
as the negative of the covariant component.  Then
\eqref{e:diraciii} simplifies to
\begin{multline}\label{e:diracvi}
   \bigl \{ \,
   i \, \gamma^0 \,
   \partial_{\tau}
   +
   i \,  \gamma^1 \,
   [ \, \partial_{\rho} + 1/ (2 \rho) \, ]
   +
   i \, \gamma^2 \, \partial_{\theta} / \rho
   \\
   +
   \gamma^3 \,
   \bigl [ \,
      i \, \partial_{\eta}
      +
      e \, A(\tau) \,
   \bigr ] / \tau
   -
   M \,
   \bigr \} \, \hat{\phi}(x)
   =
   0 \>,
\end{multline}
which is the equation we want to solve.  Here $\hat{\phi}(x)$
field obeys the simpler anticommutation relation
\begin{equation}\label{e:phifieldanticomm}
   \AntiComm{
      \hat{\phi}_{\alpha}^{\phantom\dagger}(\tau,\bx) }
            {
      \hat{\phi}_{\alpha'}^{\dagger}(\tau,\bx') }
   =
   \delta_{\alpha,\alpha'} \,
   \delta_{\bx,\bx'} \>,
\end{equation}
where $\bx = \Set{\rho,\theta,\eta}$.  Our notation is
explained in App.~\ref{s:notation}.

%
%
\subsection{Mode expansion}
\label{ss:modeexpansion}

An expansion of the $\hat{\phi}(x)$ field in terms of
transverse helicity eigenstates can be carried out using the
separation of variables methods explained in
Ref.~\onlinecite{ref:CooperDawsonMihaila06} and further
discussed App.~\ref{s:transhelicity}.  The expansion is given
by
\begin{equation}\label{e:phiexpansion}
   \hat{\phi}(x)
   =
   \sum_{\bk,\lambda}
   \hat{A}_{\bk}^{(\lambda)} \,
   \phi_{\bk}^{(\lambda)}(x) \>,
\end{equation}
where $\bk \defby \Set{\keta, \kperp, m, h}$ with
\begin{equation}\label{e:sumbkdef}
   \sum_{\bk}
   \defby
   \Int{\keta} \int_{0}^{+\infty}
      \frac{\kperp \, \rd \kperp}{ 2\pi } \!\!
   \sum_{m =-\infty}^{+\infty} \>
   \sum_{h = \pm 1} \>.
\end{equation}
(See App.~\ref{s:notation} for our notation.) Here $\lambda
= \pm 1$ labels initial positive and negative energy states, $h
= \pm 1$ labels the transverse helicity of the state, and $m$
the value of the $z$-component of the angular momentum
operator.  The time-dependent spinor mode functions
$\phi_{\bk}^{(\lambda)}(x)$ are given by
\begin{equation}\label{e:phimodedef}
   \phi_{\bk}^{(\lambda)}(x)
   \defby
   e^{i \keta \eta} \,
   \begin{pmatrix}
      \phi_{(+);k}^{(\lambda)}(\tau) \,
      \chi_{k_m,+h}^{\phantom{(}}(\rho,\theta)
      \\
      \phi_{(-);k}^{(\lambda)}(\tau) \,
      \chi_{k_m,-h}^{\phantom{(}}(\rho,\theta)
   \end{pmatrix} \>.
\end{equation}
where $k \defby \Set{\keta,\kperp,h}$ and $k_m = \Set{\kperp,
m}$, and where the transverse helicity eigenvectors
$\chi_{k_m,h}(\rho,\theta)$ are given by
\begin{equation}\label{e:chirhodef}
   \chi_{k_m,h}^{\phantom{(}}(\rho,\theta)
   =
   \frac{e^{i ( m + 1/2 ) \, \theta }}{\sqrt{2}}
   \begin{pmatrix}
      J_{m}(\kperp \rho)
      \\
      - h \, J_{m+1}(\kperp \rho)
   \end{pmatrix} \>.
\end{equation}
In App.~\ref{s:transhelicity} in Eqs.~\eqref{h.e:chinorm} and
\eqref{h.e:chicomp}, we show that they are orthogonal and
complete.

The $\phi_{\pm;k}^{(\lambda)}(\tau)$ mode functions form a two-dimensional spinor,
\begin{equation}\label{e:phispinor}
   \phi_{k}^{(\lambda)}(\tau)
   =
   \begin{pmatrix}
      \phi_{(+);k}^{(\lambda)}(\tau)
      \\
      \phi_{(-);k}^{(\lambda)}(\tau)
   \end{pmatrix}
\end{equation}
which satisfies the equations of motion
\begin{equation}\label{e:phipmeom}
   i \partial_{\tau} \, \phi_{k}^{(\lambda)}(\tau)
   =
   H_{k}(\tau) \,
   \phi_{k}^{(\lambda)}(\tau) \>,
\end{equation}
where the Hermitian matrix $H_{k}(\tau)$ satisfies:
\begin{equation}\label{e:Hdef}
\begin{split}
   H_k(\tau)
   &=
   \begin{pmatrix}
      + M &
      \peta(\tau) - i h \kperp
      \\
      \peta(\tau) + i h \kperp
      &
      - M
   \end{pmatrix}
   \\
   &=
   \bk_{k}(\tau) \cdot \bsigma \>,
\end{split}
\end{equation}
with $\peta(\tau) = [ \, \keta - e A(\tau) \, ] / \tau$ the \emph{kinetic} momentum, and where
\begin{equation}\label{e:kdef}
   \bk_{k}(\tau)
   =
   \peta(\tau) \, \be_{1}
   +
   h \kperp \, \be_{2}
   +
   M \, \be_{3} \>.
\end{equation}
We define a density matrix $\rho_{k}^{(\lambda)}(\tau)$ and ``polarization'' vector $\bP_{k}^{(\lambda)}(\tau)$ by
\begin{equation}\label{e:rhobpdef}
   \rho_{k}^{(\lambda)}(\tau)
   =
   \phi_{k}^{(\lambda)}(\tau) \,
   \phi_{k}^{(\lambda)\dagger}(\tau)
   =
   \frac{1}{2} \,
   [ \, 1 + \bP_{k}^{(\lambda)}(\tau) \cdot \bsigma \, ] \>,
\end{equation}
so that from Eq.~\eqref{e:phipmeom}, the polarization vector satisfies:
\begin{equation}\label{e:eompol}
   \partial_{\tau} \bP_{k}^{(\lambda)}(\tau)
   =
   2 \, \bk_{k}(\tau) \times \bP_{k}^{(\lambda)}(\tau) \>.
\end{equation}
We find an adiabatic expansion to second order of the
polarization vector in App.~\ref{s:adiabatic}.  Since
$H_k(\tau)$ in \eqref{e:Hdef} is Hermitian, the length of the
spinors $\phi_{k}^{(\lambda)}(\tau)$ is conserved
\begin{equation}\label{e:phicons}
   \partial_{\tau} \,
   \bigl [ \,
      \phi_{k}^{(\lambda)\dagger}(\tau) \,
      \phi_{k}^{(\lambda')}(\tau) \,
   \bigr ]
   =
   0 \>.
\end{equation}
So if we choose the two spinors labeled by $\lambda$ to be
orthogonal at $\tau = \tau_0$, then they remain orthogonal for
all~$\tau$.  In Sec.~\ref{ss:initial} below we do this, so we
can assume that these spinors are orthogonal and complete for
all values of $\tau$
\begin{subequations}\label{e:phiorthocomp}
\begin{align}
   \phi_{k}^{(\lambda)\,\dagger}(\tau) \,
   \phi_{k}^{(\lambda')}(\tau)
   &=
   \delta_{\lambda,\lambda'} \>,
   \label{e:phiortho} \\
   \sum_{\lambda=\pm}
   \phi_{k}^{(\lambda)}(\tau) \,
   \phi_{k}^{(\lambda)\,\dagger}(\tau)
   &=
   1 \>.
   \label{e:phicomplete}
\end{align}
\end{subequations}
Probability conservation also requires that the polarization
vector $\bP_{k}^{(\lambda)}(\tau)$ for both of these solutions
to remain on the unit sphere for all time $\tau$.

Using the orthogonal relations \eqref{h.e:chinorm} and
\eqref{e:phiortho}, we can invert expansion
\eqref{e:phiexpansion} to obtain for any time $\tau$,
\begin{equation}\label{e:invertphiexp}
   \hat{A}_{\bk}^{(\lambda)}
   =
   \sum_{\mathbf{x}}
   \phi_{\bk}^{(\lambda)\dagger}(x) \,
   \hat{\phi}(x) \>,
\end{equation}
where our notation is explained in App.~\ref{s:notation}.
Using \eqref{e:phifieldanticomm}, the mode operators obey the anticommutation relation
\begin{equation}\label{e:Aanticomm}
   \AntiComm{ \hat{A}_{\bk}^{(\lambda)} }
            { \hat{A}_{\bk'}^{(\lambda')\,\dagger} }
   =
   \delta_{\lambda,\lambda'}^{\phantom(} \,
   \delta_{\bk,\bk'}^{\phantom(} \>.
\end{equation}
It is traditional to define separate positive and negative energy mode operators by setting
\begin{equation}\label{e:tradition}
   \hat{A}_{\bk}^{(+)}
   =
   \hat{a}_{\bk}^{\phantom(} \>,
   \qquad\text{and}\qquad
   \hat{A}_{\bk}^{(-)}
   =
   \hat{b}_{-\bk}^{\dagger} \>.
\end{equation}
We choose our initial state to be the vacuum with no particles or anti-particles present.  Then
\begin{equation}\label{e:abvac}
   \hat{a}_{\bk}^{\phantom(} \, \Ket{0} = 0 \>,
   \qquad\text{and}\qquad
   \hat{b}_{\bk}^{\phantom(} \, \Ket{0} = 0 \>.
\end{equation}
This means that
\begin{equation}\label{e:expectcommAA}
   \Expect{ \Comm{ \hat{A}_{\bk}^{(\lambda)\,\dagger} }
                 { \hat{A}_{\bk'}^{(\lambda')} } }
   =
   -
   \lambda \,
   \delta_{\lambda,\lambda'}^{\phantom(} \,
   \delta_{\bk,\bk'}^{\phantom(} \>,
\end{equation}
a result we will use in Sec.~\ref{ss:maxwell} below.

%
%
\subsection{Initial conditions}
\label{ss:initial}

There have been several methods used to set initial conditions
for the fermion field.  We investigated two of these methods in
Ref.~\onlinecite{Mihaila:2008dp} and came to the conclusion
that both methods produce essentially the same results, so we
choose the simpler ``one-field'' method here.

Near $\tau = \tau_0 \defby 1/M$ where we take $A(\tau_0) = 0$,
the Hamiltonian \eqref{e:Hdef} is approximately independent of
$\tau$, $H_k(\tau) \approx H_{0;k}$, where
\begin{equation}\label{e:H0def}
   H_{0;k}
   =
   M
   \begin{pmatrix}
      + 1 &
      \keta - i h \barkperp
      \\
      \keta + i h \barkperp
      &
      - 1
   \end{pmatrix}
   =
   M \, \bk_{0;k} \cdot \bsigma \>,
\end{equation}
where
\begin{equation}\label{e:k0def}
   \bk_{0;k}
   =
   \keta \, \be_{1}
   +
   h \barkperp \, \be_{2}
   +
   \be_{3} \>,
\end{equation}
with $\barkperp = \kperp / M$.  We write the eigenvalue equation for the Hamiltonian $H_{0;k}$ as
\begin{gather}\label{e:eigenvalueH0}
   H_{0;k} \, \phi_{0;k}^{(\lambda)}
   =
   \omega_{0;k}^{(\lambda)} \, \phi_{0;k}^{(\lambda)} \>,
   \\
   \text{with} \quad
   \omega_{0;k}^{(\lambda)} = \lambda M \omega_{0;k} \>,
   \quad
   \omega_{0;k} = \sqrt{ \keta^2 + \barkperp^2 + 1 } \>,
   \notag
\end{gather}
and where
\begin{subequations}\label{e:phi0pmdefs}
\begin{align}
   \phi_{0;k}^{(+)}
   &=
   \sqrt{ \frac{\omega_{0;k} + 1}{2\, \omega_{0;k}} }
   \begin{pmatrix}
      1
      \\
      \zeta_k
   \end{pmatrix} \>,
   \label{e:phi0pdef} \\
   \phi_{0;k}^{(-)}
   &=
   \sqrt{ \frac{\omega_{0;k} + 1}{2\, \omega_{0;k}} }
   \begin{pmatrix}
      - \zeta_k^{\ast}
      \\
      1
   \end{pmatrix} \>,
   \label{e:phi0mdef}
\end{align}
\end{subequations}
with $\zeta_k = ( \keta + i h \barkperp ) / ( \omega_{0;k} + 1
)$.  We use these eigenvalues for initial values of the spinors
$\phi_{k}^{(\lambda)}(\tau)$ at $\tau = \tau_0$,
\begin{equation}\label{e:phiiitial}
   \phi_{k}^{(\lambda)}(\tau_0)
   =
   \phi_{0;k}^{(\lambda)} \>,
\end{equation}
which defines what we call positive and negative energy
solutions of the full Dirac equation.  Since the initial
spinors are orthogonal and complete, the full solutions of the
Dirac are also orthogonal and complete.  The density matrix
$\rho_{0;k}^{(\lambda)}$ at $\tau_0$ is given by
\begin{equation}\label{e:rho0bp0def}
   \rho_{0;k}^{(\lambda)}
   =
   \phi_{0;k}^{(\lambda)} \,
   \phi_{0;k}^{(\lambda)\dagger}
   =
   \frac{1}{2} \,
   [ \, 1 + \bP_{0;k}^{(\lambda)} \cdot \bsigma \, ] \>,
\end{equation}
where the initial polarization vector $\bP_{0;k}^{(\lambda)}$ is given by
\begin{equation}\label{e:bp0def}
   \bP_{0;k}^{(\lambda)}
   =
   \lambda \, \bk_{0;k} / \omega_{0;k} \>.
\end{equation}

%
%
\subsection{Maxwell's equation}
\label{ss:maxwell}

In boost-invariant coordinates, Maxwell's equation reads:
\begin{equation}\label{me.e:maxwellcov}
   \frac{1}{\sqrt{-g}} \,
   \partial_{\mu}
   \bigl [ \,
      \sqrt{-g} \, F^{\mu\nu}(x) \,
   \bigr ]
   =
   J^{\nu}(x) \>,
\end{equation}
where $\sqrt{-g} = \rho \tau$.  Now $A_{\mu}(x) = ( \, 0, 0, 0,
- A(\tau) \, )$, so the only non-vanishing elements of the
field tensor are:
\begin{equation}\label{me.e:Edef}
   F_{\tau,\eta}(x)
   =
   - F_{\eta,\tau}(x)
   =
   -
   \partial_\tau A(\tau)
   \defby
   \tau \, E(\tau) \,
\end{equation}
This last equation defines what we call the electric field
$E(\tau) \defby - \, \partial_\tau A(\tau) \, / \, \tau$.  Then
using the metric $g^{\mu\nu}(x) = \Diag{ 1, -1, - 1/ \rho,
-1/\tau^2 }$, we get:
\begin{equation}\label{me.e:Fupper}
   F^{\tau,\eta}(\tau)
   =
   - F^{\eta,\tau}(\tau)
   =
   -
   E(\tau) / \tau \>,
\end{equation}
and Maxwell's equation becomes:
\begin{equation}\label{me.e:maxwellcovII}
   \partial_\tau E(\tau)
   =
   - J(\tau) \>.
\end{equation}
Here we have defined a ``reduced'' current $J(\tau)$ by:
\begin{equation}\label{me.e:jsdef}
\begin{split}
   J(\tau)
   &=
   \frac{e \, \tau}{2} \,
   \Expect{
      \Comm{ \hat{\bar{\psi}}(x) }
           { \tilde{\gamma}^{\eta}(\tau) \,
             \hat{\psi}(x) } }
   \\
   &=
   \frac{e}{2 \, \tau} \,
   \Expect{
      \Comm{ \hat{\phi}^{\dagger}(x) }
      { \gamma^0 \gamma^3 \, \hat{\phi}(x) } } \>,
\end{split}
\end{equation}
Using the field expansion \eqref{e:phiexpansion} and the
expectation value \eqref{e:expectcommAA} of the mode operators,
we find for the reduced current:
\begin{align}
   J(\tau)
   &=
   \frac{e}{2 \tau}
   \sum_{\mathbf{k},\mathbf{k}'} \sum_{\lambda,\lambda'}
   \bigl [ \,
      \phi_{\bk}^{(\lambda)\,\dagger}(x) \,
      \gamma^0 \gamma^3 \,
      \phi_{\bk'}^{(\lambda')\phantom\dagger}(x) \,
   \bigr ] \,
   \notag \\[-7pt]
   & \qquad\qquad\qquad \times
   \Expect{ \Comm{ \hat{A}_{\bk}^{(\lambda)\,\dagger} }
                 { \hat{A}_{\bk'}^{(\lambda')} } }
   \notag \\
   &=
   -
   \frac{e}{2 \tau}
   \sum_{\mathbf{k},\lambda}
   \lambda \,
   \bigl [ \,
      \phi_{\bk}^{(\lambda)\,\dagger}(x) \,
      \gamma^0 \gamma^3 \,
      \phi_{\bk}^{(\lambda)\phantom\dagger}(x) \,
   \bigr ]
   \label{me.e:js}
\end{align}
Now since
\begin{equation}\label{e:gamma0gamma3}
   \gamma^0 \gamma^3
   =
   \begin{pmatrix}
      0 & \sigma_3 \\
      \sigma_3 & 0
   \end{pmatrix} \>,
\end{equation}
and using the fact that $\sigma_3 \,
\chi_{k_m,h}^{\phantom{(}}(\rho,\theta) =
\chi_{k_m,-h}^{\phantom{(}}(\rho,\theta)$, and the relation
\begin{multline}\label{e:sumchichi}
   \sum_{m=-\infty}^{+\infty}
   \chi_{k_m,h}^{\dagger}(\rho,\theta) \,
   \chi_{k_m,h}^{\phantom{(}}(\rho,\theta)
   \\
   =
   \frac{1}{2}
   \sum_{m=-\infty}^{+\infty}
   \bigl [ \,
      J^2_{m}(\kperp \rho)
      +
      J^2_{m+1}(\kperp \rho) \,
   \bigr ]
   =
   1 \>,
\end{multline}
we find from \eqref{e:phimodedef} that the reduced current can be written as
\begin{align}
   J(\tau)
   &=
   -
   \frac{e}{2 \tau}
   \sum_{k,\lambda}
   \lambda \,
   \Tr{ \rho_{k}^{(\lambda)}(\tau) \, \sigma_1 }
   \label{e:Jtau} \\
   &=
   -
   \frac{e}{\tau}
   \sum_{k}
   P_{1;\kperp,h}^{(+)}(\peta,\tau)
   =
   -
   e
   \sum_{p}
   P_{1;\kperp,h}^{(+)}(\peta,\tau) \>.
   \notag
\end{align}
Here we have used the completeness statement
\eqref{e:phicomplete} to write the current in terms of positive
energy solutions only.  In the last line, we changed
integration variables from $k_{\eta}$ to $\peta$, using $\rd
\peta = \rd k_{\eta} / \tau$, and defined
$\bP_{\kperp,h}(\peta,\tau) \defby \bP_k(\tau)$.  Maxwell's
equation \eqref{me.e:maxwellcovII} becomes:
\begin{equation}\label{e:maxwellfinal}
   \partial_{\tau} E(\tau)
   =
   e
   \sum_{p}
   P_{1;\kperp,h}^{(+)}(\peta,\tau) \>.
\end{equation}
Recall that $P_{1;\kperp,h}^{(+)}(\peta,\tau)$ is the first
component of the positive energy polarization vector.
Eq.~\eqref{e:eompol} with initial condition \eqref{e:bp0def},
and Eq.~\eqref{e:maxwellfinal} need to be solved simultaneously
for the system dynamics.

As it stands, the integral for the current in
Eq.~\eqref{e:maxwellfinal} diverges.  We renormalize it using
the adiabatic expansion of solutions of the Dirac equation we
found in App.~\ref{s:adiabatic}.  Setting $\epsilon = 1$ and
substituting \eqref{ad.e:P1comp} into
Eq.~\eqref{e:maxwellfinal} gives
\begin{align}
   & \dot{E}(\tau)
   \label{e:MEaI} \\ &
   =
   e
   \sum_{p}
   \biggl [
   \frac{\peta}{\omega}
   -
   ( \, \kperp^2 + M^2 \, ) \,
   \Bigl ( \,
      \frac{1}{4} \,
      \frac{ \ddpeta }{ \omega^5 }
      -
      \frac{5}{8} \,
      \frac{ \peta \, \dpeta^2 }{ \omega^7 } \,
   \Bigr )
   +
   \dotsb
   \biggr ] \>,
   \notag
\end{align}
where here $\omega = [ \,\peta^2 + \kperp^2 + M^2 \, ]^{1/2}$.
The dot refers to a derivative with respect to $\tau$.  So the
first term vanishes by symmetric integration over $\peta$.  For
the other terms, we note that
\begin{subequations}\label{e:pkdotsdef}
\begin{align}
   \peta(\tau)
   &=
   [ \, \keta - e A(\tau) \, ] / \tau \>,
   \label{e:pkv} \\
   \dpeta(\tau)
   &=
   -
   \peta(\tau)/\tau
   +
   e E(\tau) \>,
   \label{e:dpkv} \\
   \ddpeta(\tau)
   &=
   2 \peta(\tau)/\tau^2
   -
   e E(\tau)/\tau
   +
   e \dot{E}(\tau) \>.
\end{align}
\end{subequations}
So the only terms which survive in \eqref{e:MEaI} are
\begin{align}
   &
   \dot{E}(\tau)
   =
   2 e
   \int_{0}^{\Lambda} \!
   \frac{\kperp \, ( \, \kperp^2 + M^2 \, ) \, \rd\kperp}{2\pi} \,
   \int_{-\infty}^{+\infty} \! \frac{ \rd\peta}{2\pi} \,
   \label{e:MEaII} \\
   & \qquad \times
   \biggl [
      \frac{ e E(\tau) }{\tau} \,
      \Bigl ( \,
         -
         \frac{1}{4 \, \omega^5}
         +
         \frac{5 \, \peta^2}{4 \, \omega^7} \,
      \Bigr )
      -
      \frac{ e \dot{E}(\tau) }{ 4 \, \omega^5 }
      +
      \dotsb
   \biggr ]
   \>.
   \notag
\end{align}
Here we have introduced a cutoff $\Lambda$ in the $\kperp$
integral.  Carrying out the integrals in \eqref{e:MEaII} and
moving terms proportional to $\dot{E}(\tau)$ to the
left-hand-side, we find the adiabatic expansion of Maxwell's
equation to be
\begin{equation}\label{e:adMaxwellI}
   \bigl [ \, 1 + e^2 ( \delta e^2 ) \, \bigr ] \, \dot{E}(\tau)
   =
   e \, R[ e \, A(\tau) ] \>,
\end{equation}
where $\delta e^2$ is given by
\begin{equation}\label{e:deltae2def}
   \delta e^2
   =
   \frac{1}{6\pi^2} \, \Ln{ \Lambda/ M } \>,
\end{equation}
and $R[ e \, A(\tau) ]$ is a finite functional of the product
$e A(\tau)$, or derivatives of this quantity.  We define the
renormalized charge $e_r$ by
\begin{equation}\label{e:eRdef}
   e_r^2
   =
   \frac{e^2}{1 + e^2 ( \delta e^2 ) } \>.
\end{equation}
Then since $e A(\tau) = e_r A_r(\tau)$, the adiabatic expansion
of Maxwell's equation \eqref{e:adMaxwellI} reduces to
\begin{equation}\label{e:adMaxwellII}
   \dot{E}_r(\tau)
   =
   e_r \, R[ e_r \, A_r(\tau) ] \>,
\end{equation}
which is now finite.  We conclude that we can regularize
Maxwell's equation by subtracting from the integrand the
adiabatic expansion of $P_{1;\kperp,h}^{(+)}(\peta,\tau)$ and
in addition renormalizing the charge.  This gives the equation
\begin{align}
   &\partial_{\tau} E(\tau)
   =
   \frac{e}{1 + e^2 ( \delta e^2 ) }
   \sum_{p}^{\Lambda}
   \biggl [ \,
      P_{1;\kperp,h}^{(+)}(\peta,\tau)
      \label{e:maxwellsub} \\ & \qquad
      -
      \frac{\peta}{\omega}
      +
      \frac{ e E(\tau) }{\tau} \,
      ( \, \kperp^2 + M^2 \, ) \,
      \Bigl ( \,
         \frac{1}{4 \, \omega^5}
         -
         \frac{5 \, \peta^2}{4 \, \omega^7} \,
      \Bigr )
   \biggr ] \>.
   \notag
\end{align}

%
%

\section{Energy-momentum tensor}
\label{s:energymomentum}

In the boost-invariant coordinate system, the average value of
the total energy-momentum tensor is given by Eqs.~(4.1) and
(4.2) of Ref.~\onlinecite{r:Cooper:1993uq}, and is the sum of
two terms (notice sign convention)
\begin{equation}\label{em.e:Tmunu}
\begin{split}
   T_{\mu\nu}
   &=
   T_{\mu\nu}^{\text{matter}}
   +
   T_{\mu\nu}^{\text{field}}
   \\
   &=
   \Diag{
      \calE,
      \calPperp,
      \rho^2 \calPtheta,
      \tau^2 \calPeta } \>,
\end{split}
\end{equation}
where the matter and field contributions are given by
\begin{subequations}\label{em.e:Tmf}
\begin{align}
   T_{\mu\nu}^{\text{mat}}
   &=
   \frac{1}{4}
   \bigl \langle
      \Comm{ \hat{\bar{\psi}}(x) }
           { \tilde{\gamma}_{(\mu}^{\phantom\ast}(x) \,
             ( i \, D_{\nu)}^{\phantom\ast} \, \hat{\psi}(x) ) }
      +
      \text{h.c.} \,
   \bigr \rangle
   \label{em.e:Tmatter} \\
   T_{\mu\nu}^{\text{field}}
   &=
   g_{\mu\nu} \, \frac{1}{4} \, F^{\alpha\beta} F_{\alpha\beta}
   +
   F_{\mu\alpha} \, g^{\alpha\beta} \,F_{\beta\nu} \>.
   \label{em.e:Tfield}
\end{align}
\end{subequations}
Here $D_{\mu} = \partial_{\mu} + i e \, A_{\mu}(x)$ and the
subscript notation $(\mu,\nu)$ means to symmetrize the term.
From our results for the field tensor in  Eq.~\eqref{me.e:Edef}
in Sec.~\ref{ss:maxwell}, the field part of the energy-momentum
tensor is given by
\begin{equation}\label{em.e:fieldEMtensor}
   T_{\mu\nu}^{\text{field}}
   =
   \frac{1}{2} \,
   \Diag{ E^2, E^2, \rho^2 E^2, -\tau^2 E^2 } \>.
\end{equation}
We denote the matter part of the energy-momentum tensor as:
\begin{equation}\label{matter:Tmunu}
   T_{\mu\nu}^{\text{matter}}
   =
   \Diag{ \varepsilon, p_\perp, \rho^2 \, p_\theta, \tau^2 \, p_\eta } \>.
\end{equation}
Because of the conventions adapted in Eq.~\eqref{em.e:Tmunu},
the total energy and pressures are obtained by adding a factor
of $\pm E^2/2$ to the matter terms.

For the matter field, we first note that $D_{\mu} \,
\hat{\psi}(x) = S(x) \, \nabla_{\mu} \, \hat{\phi}(x) /
\sqrt{\tau}$, where $\nabla_{\mu} = \partial_{\mu} +
\Pi_{\mu}(x) + i e A_{\mu}(x)$ is the covariant derivative
defined in Eq.~\eqref{e:defcovder}.  So using the use the
notation in App.~\ref{s:notation}, the field expansion
\eqref{e:phiexpansion}, and the expectation value
\eqref{e:expectcommAA} of the mode operators, the diagonal
components of the matter energy-momentum tensor
\eqref{em.e:Tmatter} are given by (no sum over $\mu$)
\begin{widetext}
\begin{align}
   & T_{\mu\mu}^{\text{mat}}
   =
   \frac{1}{4}
   \biggl \langle
   \Bigl [ \,
      \Bigl ( \,
         \frac{\hat{\phi}^{\dagger}(x)}{\sqrt{\tau}} \,
      \Bigr ) \,
      \gamma^0 \, , \,
      \bar{\gamma}_{\mu}^{\phantom\ast}(x) \,
      \Bigl \{ \,
      i \nabla_{\mu}
      \Bigl ( \,
         \frac{\hat{\phi}(x)}{\sqrt{\tau}} \,
      \Bigr ) \,
      \Bigr \}
   \Bigr ]
   +
   \Bigl [ \,
      \Bigl \{ \,
      i \bar{\nabla}_{\mu}
      \Bigl ( \,
         \frac{\hat{\phi}(x)}{\sqrt{\tau}} \,
      \Bigr ) \,
      \Bigr \}^{\dagger} \,
      \gamma^0 \, , \,
      \bar{\gamma}_{\mu}^{\phantom\ast}(x) \,
      \Bigl ( \,
         \frac{\hat{\phi}(x)}{\sqrt{\tau}} \,
      \Bigr ) \,
   \Bigr ]
   \biggr \rangle
   \label{em.e:TmatterII} \\
   &=
   \frac{1}{4}
   \sum_{\bk,\bk'} \sum_{\lambda,\lambda'}
   \biggl [ \,
      \Bigl ( \,
         \frac{\hat{\phi}_{\bk}^{(\lambda)\dagger}(x)}
              {\sqrt{\tau}} \,
      \Bigr ) \,
      \gamma^0 \,
      \bar{\gamma}_{\mu}^{\phantom\ast}(x) \,
      \Bigl \{ \,
      i \nabla_{\mu}
      \Bigl ( \,
         \frac{\hat{\phi}_{\bk'}^{(\lambda')}(x)}
              {\sqrt{\tau}} \,
      \Bigr ) \,
      \Bigr \}
      +
      \Bigl \{ \,
      i \bar{\nabla}_{\mu}
      \Bigl ( \,
         \frac{\hat{\phi}_{\bk}^{(\lambda)}(x)}
              {\sqrt{\tau}} \,
      \Bigr ) \,
      \Bigr \}^{\dagger} \,
      \gamma^0 \,
      \bar{\gamma}_{\mu}^{\phantom\ast}(x) \,
      \Bigl ( \,
         \frac{\hat{\phi}_{\bk'}^{(\lambda')}(x)}
              {\sqrt{\tau}} \,
      \Bigr ) \,
   \biggr ] \,
   \Expect{ \Comm{ \hat{A}_{\bk}^{(\lambda)\,\dagger} }
                 { \hat{A}_{\bk'}^{(\lambda')} } }
   \notag \\
   &=
   \frac{-1}{4}
   \sum_{\bk,\lambda}
   \lambda \,
   \biggl [ \,
      \Bigl ( \,
         \frac{\hat{\phi}_{\bk}^{(\lambda)\dagger}(x)}
              {\sqrt{\tau}} \,
      \Bigr ) \,
      \Bigl \{ \,
      i
      \gamma^0 \,
      \bar{\gamma}_{\mu}^{\phantom\ast}(x) \,
      \nabla_{\mu}
      \Bigl ( \,
         \frac{\hat{\phi}_{\bk}^{(\lambda)}(x)}
              {\sqrt{\tau}} \,
      \Bigr ) \,
      \Bigr \}
      +
      \Bigl \{ \,
      i \gamma^0 \,
      \bar{\gamma}_{\mu}^{\phantom\ast}(x) \,
      \bar{\nabla}_{\mu}
      \Bigl ( \,
         \frac{\hat{\phi}_{\bk}^{(\lambda)}(x)}
              {\sqrt{\tau}} \,
      \Bigr ) \,
      \Bigr \}^{\dagger} \,
      \Bigl ( \,
         \frac{\hat{\phi}_{\bk}^{(\lambda)}(x)}
              {\sqrt{\tau}} \,
      \Bigr ) \,
   \biggr ] \>,
   \notag
\end{align}
\end{widetext}
where we used the fact that $\bar{\gamma}_{\mu}(x)$
\emph{anti}commutes with $\Pi_{\mu}(x)$, and the relation
$\gamma^0 \Pi_{\mu}^{\dagger}(x) \gamma^0 = - \Pi_{\mu}(x)$.
Here we have defined the covariant derivatives
\begin{equation}\label{em.e:nnbar}
\begin{split}
   \nabla_{\mu}^{\phantom\ast}
   &=
   \partial_{\mu}
   +
   \Pi_{\mu}(x)
   +
   i e \, A_{\mu}(x) \>,
   \\
   \bar{\nabla}_{\mu}^{\phantom\ast}
   &=
   \partial_{\mu}
   -
   \Pi_{\mu}(x)
   +
   i e \, A_{\mu}(x) \>.
\end{split}
\end{equation}

%
%

\newpage

\subsection{Energy Density }
\label{ss:energy}

For matter energy density term, $\nabla_{0} = \partial_\tau$
and $\bar{\gamma}_{\tau} = \bar{\gamma}^{\tau} = \gamma^0$, so
using \eqref{em.e:TmatterII}, we find
\begin{align}
   \varepsilon(\tau)
   &=
   - \frac{i}{4 \tau}
   \sum_{\bk,\lambda}
   \lambda \,
   \bigl [ \,
      \phi_{\bk}^{(\lambda)\,\dagger}(x) \,
      \overleftrightarrow{\partial_{\tau}} \,
      \phi_{\bk}^{(\lambda)\phantom\dagger}(x) \,
   \bigr ]
   \notag \\
   &=
   -
   \frac{1}{2 \tau}
   \sum_{k,\lambda}
   \Tr{ \rho_{\bk}^{(\lambda)}(\tau) \, H_{\bk}(\tau) }
   \label{em.e:Ttt}
   \\ \notag
   &=
   -
   \sum_{p} \>
   \bk_{\kperp,h}^{\phantom{(}}(\peta) \cdot
   \bP^{(+)}_{\kperp,h}(\peta,\tau) \>.
\end{align}
So from \eqref{em.e:fieldEMtensor}, the total energy density is given by
\begin{equation}\label{em.e:Entot}
   \calE
   =
   -
   \sum_{p} \>
   \bk_{\kperp,h}^{\phantom{(}}(\peta) \cdot
   \bP^{(+)}_{\kperp,h}(\peta,\tau)
   +
   \frac{E^2}{2} \>.
\end{equation}
As it stands, the integral for the energy density  in
Eq.~\eqref{em.e:Entot} diverges.  We find the form of theses
divergences by substituting the adiabatic expansion given in
Eq.~\eqref{pv.e:psecondorder} and introducing a cutoff
$\Lambda_\eta$ in the $\peta$ integral.  This gives a adiabatic
approximation to the energy density of
\begin{equation*}
   \calE_a
   =
   \frac{E^2}{2}
   -
   \sum_{p}^{\Lambda}
   \biggl [ \,
      \omega
      -
      \frac{( \kperp^2 + M^2 )}{8} \,
      \frac{\dpeta^2}{\omega^5}
      \dotsb
   \biggr ] \,.
\end{equation*}
Using Eqs.~\eqref{e:pkdotsdef}, due to the symmetric
integration over $\peta$, the only terms that survive are
\begin{align}
   &
   \calE_a
   =
   \frac{E^2}{2}
   -
   \sum_{p}^{\Lambda}
   \Bigl [
      \omega
      -
      \frac{\kperp^2 + M^2}{8}
      \Bigl (
         \frac{1}{\tau^2} \frac{\peta^2}{\omega^5}
         +
         \frac{e^2 E^2}{\omega^5}
      \Bigr )
      +
      \dotsb
   \Bigr ]
   \notag \\
   &=
   \bigl [ 1 + e^2 ( \delta e^2 ) \bigr ] \frac{E^2}{2}
   \! + \!
   \frac{1}{12} \Bigl ( \frac{\Lambda}{2\pi \tau} \Bigr )^2
   \!\! -
   \sum_{p}^{\Lambda}
   \bigl (
      \omega
      +
      \dotsb
   \bigr ) \>.
   \label{em.e:adenergy}
\end{align}
The first term renormalizes the field
\begin{equation}\label{e:E2renorm}
   \bigl [ \, 1 + e^2 ( \delta e^2 ) \, \bigr ] \, \frac{E^2}{2}
   =
   \frac{e^2 E^2}{2 \er^2}
   =
   \frac{\Er^2}{2} \>.
\end{equation}
The second term contributes to the cosmological constant, as we
will see later.  The third term is related to the zero-point
energy of pairs of fermions.  We regularize the energy density
by computing the difference between $\calE$ and $\calE_a$,
$\calE^{\text{sub}} = \calE - \calE_a$, which is now finite.

%
%
\subsection{Transverse pressure}
\label{ss:tpressure}

For the matter transverse pressure term, we have
$\bar{\gamma}_{\rho} = - \bar{\gamma}^{\rho} = - \gamma^{1}$,
and $\nabla_{\rho} = \partial_{\rho}$, so from
\eqref{em.e:TmatterII}, we find:
\begin{align}
   &p_\perp(\tau)
   \label{em.e:calPperp}
   =
   - \frac{i}{4 \tau}
   \sum_{\bk,\lambda}
   \lambda
   \bigl [
      \phi_{\bk}^{(\lambda)\,\dagger}(x)
      \gamma^0 \gamma^1
      \overleftrightarrow{\partial_{\rho}}
      \phi_{\bk}^{(\lambda)\phantom\dagger}(x)
   \bigr ] \>.
\end{align}
So here we will need to find
\begin{multline*}
   \sum_{m=-\infty}^{+\infty}
   \chi_{\kperp,m,h}^{\dagger}(\rho) \,
   \sigma_x \,
   \overleftrightarrow{\partial_{\rho}} \,
   \chi_{\kperp,m,-h}^{\phantom\dagger}(\rho)
   \\
   =
   h \!
   \sum_{m=-\infty}^{+\infty}
      J_{m}(\kperp \rho) \,
      \overleftrightarrow{\partial_{\rho}} \,
      J_{m+1}(\kperp \rho)
   =
   h \kperp \>,
\end{multline*}
where we have used the relation
\begin{equation*}
   2 \, J'_m(z)
   =
   J_{m-1}(z) - J_{m+1}(z) \>.
\end{equation*}
Then \eqref{em.e:calPperp} becomes
\begin{align}
   p_\perp(\tau)
   &=
   \frac{1}{\tau}
   \sum_{k} \,
   \frac{ h \kperp}{2} \,
   \bigl [ \,
      \phi_{k}^{(+)\dagger}(\tau) \,
      \sigma_y \,
      \phi_{k}^{(+)}(\tau) \,
   \bigr ]
   \notag \\
   &=
   \sum_{p} \,
   \frac{ h \kperp}{2} \,
   P_{2;\kperp,h}^{(+)}(\peta,\tau) \>,
   \label{em.e:TperpmI}
\end{align}
and from \eqref{em.e:fieldEMtensor}, the total transverse pressure is given by
\begin{equation}\label{em.e:calPperptot}
   \calPperp
   =
   \sum_{p} \,
   \frac{ h \kperp}{2} \,
   P_{2;\kperp,h}^{(+)}(\peta,\tau)
   +
   \frac{E^2}{2} \>.
\end{equation}
In order to study the divergences in the transverse pressure,
we substitute the adiabatic expansion \eqref{ad.e:P2comp} into
\eqref{em.e:calPperptot}.  This gives
\begin{align}
   \mathcal{P}_{a; \perp}
   &=
   \frac{E^2}{2}
   +
   \sum_{p}^{\Lambda} \,
   \biggl [
      \frac{\kperp^2}{2 \omega}
      -
      \frac{ h \kperp \, M \, \dpeta}{4 \, \omega^3}
      \label{em.e:calPperpad} \\ & \qquad
      +
      \frac{\kperp^2}{2} \,
      \Bigl ( \,
         -
         \frac{1}{8} \,
         \frac{\dpeta^2}{\omega^5}
         +
         \frac{1}{4} \,
         \frac{\peta \, \ddpeta}{\omega^5}
         -
         \frac{5}{8} \,
         \frac{\peta^2 \, \dpeta^2}{\omega^7} \,
      \Bigr )
      +
      \dotsb
   \biggr ] \>.
   \notag
\end{align}
The second term in the above sum over $p$ is odd in $h$ and
therefore vanishes. From Eqs.~\eqref{e:pkdotsdef}, the only
terms that survive the $\peta$ integration are
\begin{align}
   \mathcal{P}_{a; \perp}
   &=
   \frac{E^2}{2}
   +
   \sum_{p}^{\Lambda} \,
   \biggl \{
      \frac{\kperp^2}{2 \omega}
      -
      \frac{\kperp^2}{2} \,
      \Bigl [ \,
         -
         \frac{1}{\tau^2} \,
         \Bigl ( \,
            \frac{3}{8} \,
            \frac{\peta^2}{\omega^5}
            -
            \frac{5}{8} \,
            \frac{\peta^4}{\omega^7} \,
         \Bigr )
         \notag \\ & \qquad
         -
         e^2 E^2 \,
         \Bigl ( \,
            \frac{1}{8} \,
            \frac{1}{\omega^5}
            +
            \frac{5}{8} \,
            \frac{\peta^2}{\omega^7} \,
         \Bigr )
      \Bigr ]
      +
      \dotsb
   \biggr \}
   \label{em.e:calPperpadII} \\
   &=
   \bigl [ \, 1 + e^2 ( \delta e^2 ) \, \bigr ] \, \frac{E^2}{2}
   +
   \sum_{p}^{\Lambda} \
   \biggl (
      \frac{\kperp^2}{2 \omega}
      +
      \dotsb
   \biggr ) \>.
   \notag
\end{align}
From \eqref{e:E2renorm}, the first term renormalizes the
electric field.  The term proportional to $1/\tau^2$ vanishes.
Again, the transverse pressure is regularized by subtracting
the adiabatic expression from the divergent one,
$\calPperp^{\text{sub}} = \calPperp - \mathcal{P}_{a; \perp}$.

%
%
\subsection{Shear Pressure}
\label{ss:shear}

For the shear pressure term, we have $\bar{\gamma}_{\theta} = -
\rho^2 \bar{\gamma}^{\theta} = - \rho \gamma^{2}$ and
$\Pi_{\theta} = \gamma^1 \gamma^2$.  The covariant derivatives
\eqref{em.e:nnbar} are given by
\begin{align*}
   \gamma^0 \gamma^2 \, \nabla_{\theta}
   &=
   \gamma^0 \gamma^2 \, \partial_{\theta}
   +
   \gamma^0 \gamma^1 / 2 \>,
   \\
   \gamma^0 \gamma^2 \, \bar{\nabla}_{\eta}
   &=
   \gamma^0 \gamma^2 \, \partial_{\theta}
   -
   \gamma^0 \gamma^1 / 2 \>,
\end{align*}
where
\begin{equation*}
   \gamma^0 \gamma^1
   =
   \begin{pmatrix} 0 & \sigma_x \\ \sigma_x & 0 \end{pmatrix} \>,
   \qquad
   \gamma^0 \gamma^2
   =
   \begin{pmatrix} 0 & \sigma_y \\ \sigma_y & 0 \end{pmatrix} \>.
\end{equation*}
So from \eqref{em.e:TmatterII}, we find
\begin{align}
   &
   p_\theta(\tau)
   =
   \frac{-i}{4 \rho \tau} \!\!
   \sum_{\bk,\lambda}
   \lambda
   \Bigl \{
      \phi_{\bk}^{(\lambda)\,\dagger}(x)
      \bigl [
         \bigl (
            \gamma^0 \gamma^2 \partial_{\theta}
            +
            \gamma^0 \gamma^1 / 2
         \bigr ) \,
         \phi_{\bk}^{(\lambda)}(x)
      \bigr ]
      \notag \\ & \quad
      +
      \bigl [ \,
         \bigl ( \,
            \gamma^0 \gamma^2 \, \partial_{\theta}
            -
            \gamma^0 \gamma^1 / 2 \,
         \bigr ) \,
         \phi_{\bk}^{(\lambda)}(x) \,
      \bigr ]^{\dagger} \,
      \phi_{\bk}^{(\lambda)}(x) \,
   \Bigr \}
   \>.
   \label{em.e:PthetamI}
\end{align}
So here we need to compute
\begin{multline*}
   \sum_{m=-\infty}^{+\infty}
   \chi_{k_m,h}^{\dagger}(\theta,\rho) \,
   \bigl ( \,
      \sigma_y \partial_{\theta} \pm \sigma_x / 2 \,
   \bigr ) \, \chi_{k_m,-h}^{\phantom{\dagger}}(\theta,\rho)
   \\
   =
   \frac{h}{2}
   \sum_{m=-\infty}^{+\infty}
   ( 2m + 1 ) \, J_{m}(\kperp \rho) \, J_{m+1}(\kperp \rho)
   =
   \frac{h \kperp \rho}{2} \>,
\end{multline*}
where we have used the relation
\begin{equation*}\label{em.e:BessrecII}
   2 m \, J_{m}(z)
   =
   z \,
   [ \, J_{m+1}(z) + J_{m-1}(z) \, ] \>.
\end{equation*}
Then \eqref{em.e:PthetamI} becomes
\begin{align}
   p_\theta(\tau)
   &=
   \frac{1}{\tau}
   \sum_{k} \,
   \frac{ h \kperp}{2} \,
   \bigl [ \,
      \phi_{k}^{(+)\dagger}(\tau) \,
      \sigma_y \,
      \phi_{k}^{(+)}(\tau) \,
   \bigr ]
   \label{em.e:TperpmII} \\
   &=
   \sum_{p}
   \Bigl ( \frac{ h \kperp}{2} \Bigr ) \,
   P_{2;\kperp,h}^{(+)}(\peta,\tau) \>,
   \notag
\end{align}
adding this to the shear pressure of the field, we find
\begin{equation}\label{em.e:calPthetatot}
   \calPtheta
   =
   \sum_{p} \,
   \frac{ h \kperp}{2} \,
   P_{2;\kperp,h}^{(+)}(\peta,\tau)
   +
   \frac{E^2}{2} \>.
\end{equation}
Note that $\calPtheta = \calPperp$.  The shear pressure is
renormalized exactly like the transverse pressure.

%
%
\subsection{Longitudinal pressure}
\label{ss:longpressure}

For the longitudinal pressure ($\mu=\eta$), $A_{\eta}(x) = -
A(\tau)$ and $\bar{\gamma}_{\eta}=-\tau^2 \bar{\gamma}^{\eta}=
-\tau \gamma^3$.  The covariant derivatives \eqref{em.e:nnbar}
are given by
\begin{align*}
   i \, \gamma^0 \gamma^3 \, \nabla_{\eta}
   &=
   \gamma^0 \gamma^3 \,
   [ \, i \partial_{\eta} + e A(\tau) \, ]
   +
   \gamma^5 / 2 \>,
   \\
   i  \, \gamma^0 \gamma^3 \, \bar{\nabla}_{\eta}
   &=
   \gamma^0 \gamma^3 \,
   [ \, i \partial_{\eta} + e A(\tau) \, ]
   -
   \gamma^5 / 2 \>,
\end{align*}
where
\begin{equation*}
   \gamma^5
   =
   i \gamma^0 \gamma^1 \gamma^2 \gamma^3
   =
   \begin{pmatrix} 0 & 1 \\ 1 & 0 \end{pmatrix} \>,
   \quad
   \gamma^0 \gamma^3
   =
   \begin{pmatrix} 0 & \sigma_z \\ \sigma_z & 0 \end{pmatrix} \>.
\end{equation*}
Here $\gamma^5$ flips the upper and lower components of the spinor, which leads to the equation
\begin{multline}\label{em.e:chiorthog}
   \sum_{m=-\infty}^{+\infty}
   \chi_{k_m,h}^{\dagger}(\rho,\theta) \,
   \chi_{k_m,-h}^{\phantom\dagger}(\rho,\theta)
   \\
   =
   \frac{1}{2}
   \sum_{m=-\infty}^{+\infty}
   \bigl [ \,
      J_{m}^2(\kperp\rho) - J_{m+1}^2(\kperp\rho) \,
   \bigr ]
   =
   0 \>.
\end{multline}
So terms proportional to $\gamma^5$ vanish.  Then from \eqref{em.e:TmatterII}, we are left with
\begin{align}
   p_\eta(\tau)
   &=
   \frac{1}{4 \tau^2}
   \sum_{\bk,\lambda}
   \lambda \,
   \Bigl \{
      \phi_{\bk}^{(\lambda)\,\dagger}(x)
      \bigl [
         \gamma^0 \gamma^3 \,
         ( \, i \partial_{\eta} + e A(\tau) \, ) \,
         \phi_{\bk}^{(\lambda)}(x)
      \bigr ]
      \notag \\ & \qquad\quad
      +
      \bigl [ \,
         \gamma^0 \gamma^3 \,
         ( \, i \partial_{\eta} + e A(\tau) \, ) \,
         \phi_{\bk}^{(\lambda)}(x) \,
      \bigr ]^{\dagger} \,
      \phi_{\bk}^{(\lambda)}(x) \,
   \Bigr \}
   \notag \\
   &=
   \frac{-1}{\tau^2}
   \sum_{k} \,
   [ \, \keta - e A(\tau) \, ] \, P_{1;k}(\tau)
   \label{em.e:PetamI} \\
   &=
   -
   \sum_{p}
   \peta \, P_{1;\keta,h}(\peta,\tau) \>.
   \notag
\end{align}
Adding the field pressure, we find for the total longitudinal pressure
\begin{equation}\label{e:calPetatot}
   \calPeta
   =
   -
   \sum_{p}
   \peta \, P_{1;\keta,h}(\peta,\tau)
   -
   \frac{E^2}{2} \>.
\end{equation}
From Eq.~\eqref{ad.e:P1comp}, the adiabatic expansion of the longitudinal pressure is given by
\begin{align}
   & \mathcal{P}_{a; \eta}
   =
   -
   \frac{E^2}{2}
      \label{em.e:calPetaad} \\ &
   -
   \sum_{p}^{\Lambda}
   \biggl [ \,
      \frac{\peta^2}{\omega}
      -
      ( \, \kperp^2 + M^2 \, ) \,
      \Bigl ( \,
         \frac{1}{4} \,
         \frac{ \peta \ddpeta }{ \omega^5 }
         -
         \frac{5}{8} \,
         \frac{ \peta^2 \, \dpeta^2 }{ \omega^7 } \,
      \Bigr )
      +
      \dotsb
   \biggr ] \>.
   \notag
\end{align}
From Eqs.~\eqref{e:pkdotsdef}, the only terms that survive the $\peta$ integration are
\begin{align}
   & \mathcal{P}_{a; \eta}
   =
   -
   \frac{E^2}{2}
   -
   \sum_{p}^{\Lambda}
   \biggl [ \,
      \frac{\peta^2}{\omega}
      +
      \frac{5}{8} \, e^2 E^2 \,
      ( \, \kperp^2 + M^2 \, ) \,
      \frac{\peta^2}{\omega^7}
      \notag \\ & \qquad \qquad
      -
      \frac{( \, \kperp^2 + M^2 \, )}{2 \, \tau^2} \,
      \Bigl ( \,
         \frac{\peta^2}{\omega^5}
         -
         \frac{5}{4} \,
         \frac{\peta^4}{\omega^7} \,
      \Bigr )
      +
      \dotsb
   \biggr ]
   \label{em.e:calPetaadII}
   \\ &=
   -
   \bigl [ 1 + e^2 ( \delta e^2 ) \bigr ] \frac{E^2}{2}
   +
   \frac{1}{12} \Bigl ( \frac{\Lambda}{2\pi \tau} \Bigr )^2
   \!\! -
   \sum_{p}^{\Lambda}
   \biggl (
      \frac{\peta^2}{\omega}
      +
      \dotsb
   \biggr ) \>.
   \notag
\end{align}
Again, the first term renormalizes the electric field and the
second term renormalizes the cosmological constant.  The finite
part of the longitudinal pressure is given by
$\calPeta^{\text{sub}} = \calPeta -\mathcal{P}_{a; \eta}$, as
before.

%
%
\subsection{Conservation equations}
\label{ss:conseqs}

The covariant derivative of the energy-momentum tensor in boost-invariant  coordinates is conserved
\begin{equation}\label{em.e:consI}
   T^{\mu\nu}{}_{;\mu}
   =
   \partial_{\mu} T^{\mu\nu}
   +
   \Gamma^{\mu}_{\mu\sigma} T^{\sigma\nu}
   +
   \Gamma^{\nu}_{\mu\sigma} T^{\mu\sigma}
   =
   0 \>.
\end{equation}
The only nonzero Christoffel symbols are given in
Eq.~\eqref{e:christoff}.  There are only two conservation
equations that result from Eq.~\eqref{em.e:consI}.  For
$\nu=\tau$, \eqref{em.e:consI} reduces to
\begin{align}
   \partial_\tau \, T^{\tau\tau}
   +
   T^{\tau\tau} / \tau
   +
   \tau \, T^{\eta\eta}
   &=
   0 \>,
   \notag \\
   \text{or} \qquad
   \partial_\tau \, ( \tau \calE )
   +
   \calPeta
   =
   0
   &\>.
   \label{em.e:ConsI}
\end{align}
Using the equation of motion \eqref{e:eompol} and Maxwell's
equation \eqref{e:maxwellfinal}, one can show that
Eq.~\eqref{em.e:ConsI} is automatically satisfied.

The conservation equation for $\nu=\rho$ amounts to a relation
between the transverse and shear pressures.  We find that
$\calPperp = \calPtheta$, which is satisfied by our expression
in Eqs.~\eqref{em.e:calPperptot} and \eqref{em.e:calPthetatot}.
For the one-dimensional boost invariant expansion we had
instead for the equation for the energy density,
\begin{equation}
   \partial_\tau \, ( \tau \calE )
   +
   \calP
   =
   0 \>.
\end{equation}

\begin{figure}[t!]
   \centering
   \includegraphics[width=0.9\columnwidth]{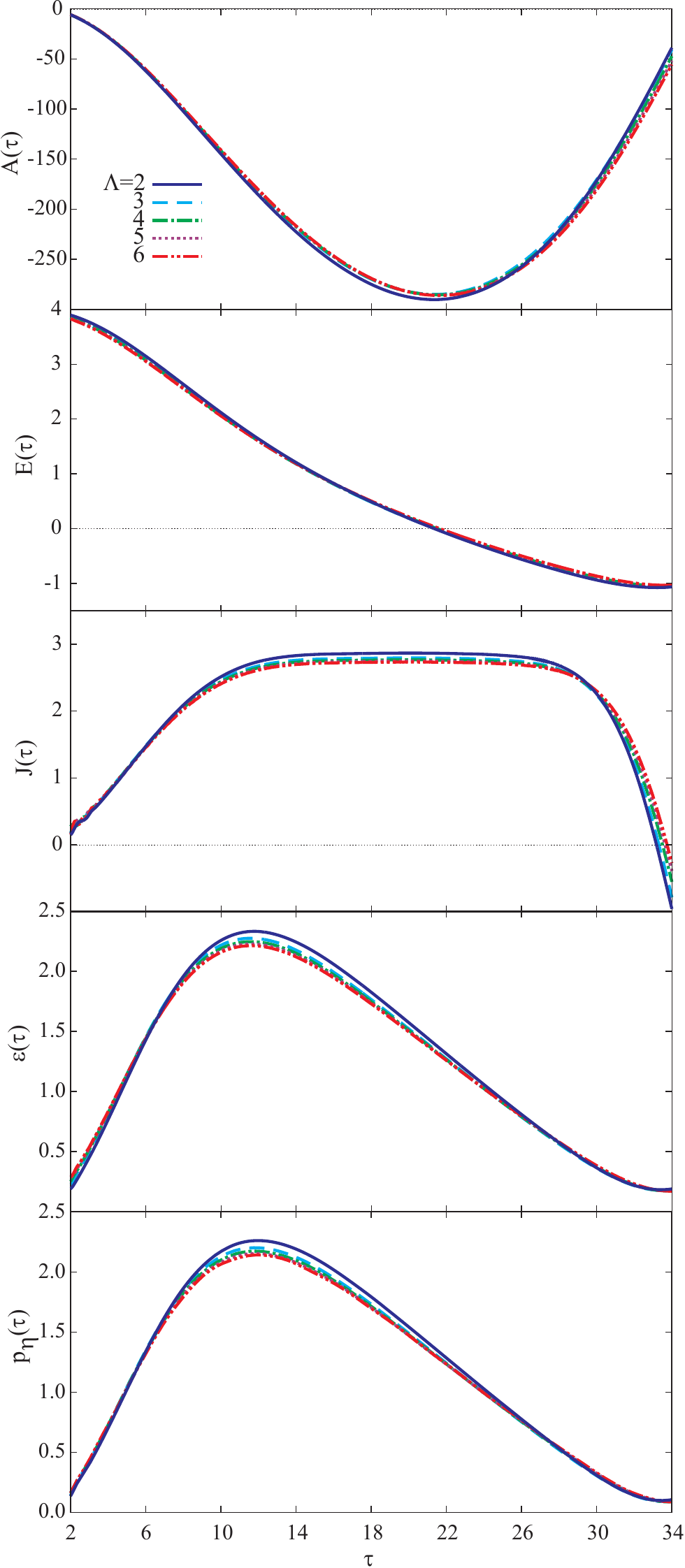}
   \caption{\label{fig1}(Color online)
   Convergence of the proper-time evolution of the electromagnetic fields, current,
   energy and longitudinal pressure with respect to the cutoff~$\Lambda$, in (3+1)-dimensional QED.
   Here we choose $m=1$, $A(\tau_0)=0$ and $E(\tau_0)=4$.}
\end{figure}

\begin{figure}[t!]
   \centering
   \includegraphics[width=0.9\columnwidth]{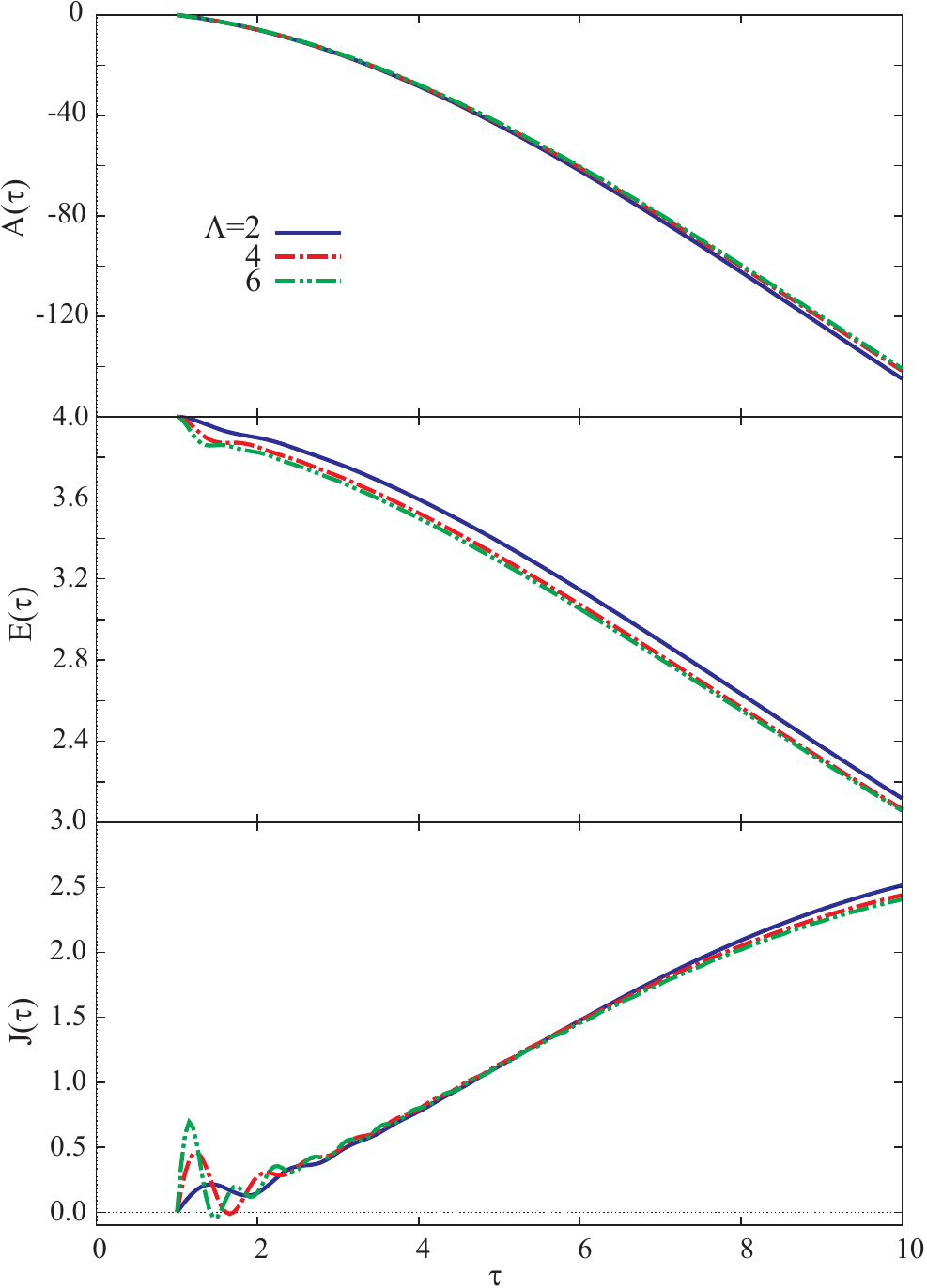}
   \caption{\label{fig2}(Color online)
   Early proper-time evolution of the electromagnetic fields and current with respect to the
   cutoff~$\Lambda$, in (3+1)-dimensional QED.
   We note that for $\tau<5$ the results are dependent on the cutoff, whereas at later proper
   times the results become insensitive to the cutoff.
   This is an artifact of our choice of initial conditions, which are not consistent at early times with the
   adiabatic-expansion-based substraction scheme.}
\end{figure}

\begin{figure}[t!]
   \centering
   \includegraphics[width=0.9\columnwidth]{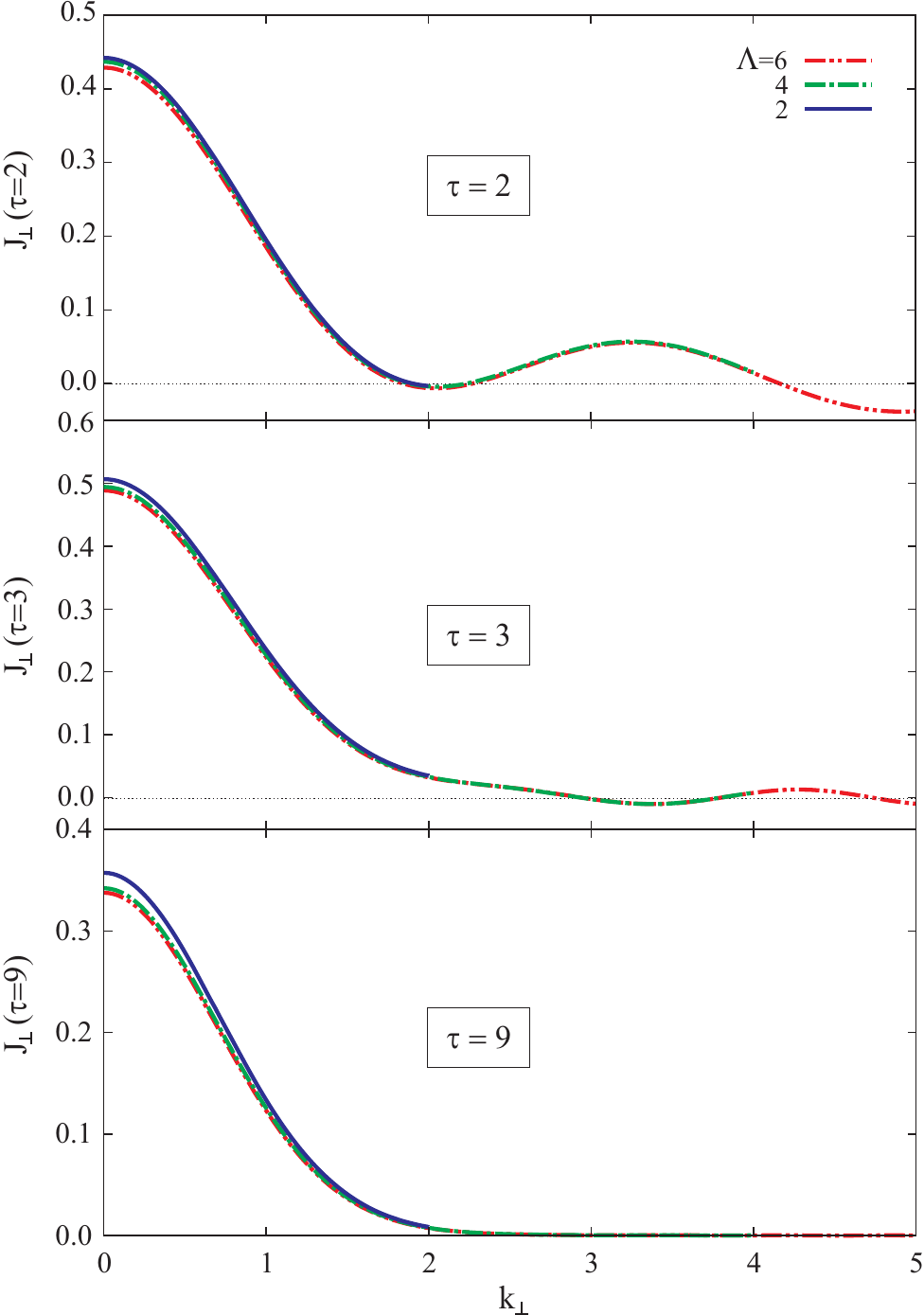}
   \caption{\label{fig3}(Color online)
   Proper-time evolution of the transverse ($k_\perp$-projected) distribution of the
   current for $\tau$ values of 2, 3 and~9.
   The oscillations of the transverse distributions of the current present at early proper times
   are consequences of the inconsistency between the adiabatic-expansion regularization scheme and our
   choice of initial conditions which are designed to keep the storage and time requirements  of the simulation to a minimum.
   At later proper times, these oscillations dampen out and eventually disappear.
   Therefore, the early proper-time evolution should be regarded as ``unphysical'' and will be disregarded.
   }
\end{figure}

\begin{figure}[t!]
   \centering
   \includegraphics[width=0.9\columnwidth]{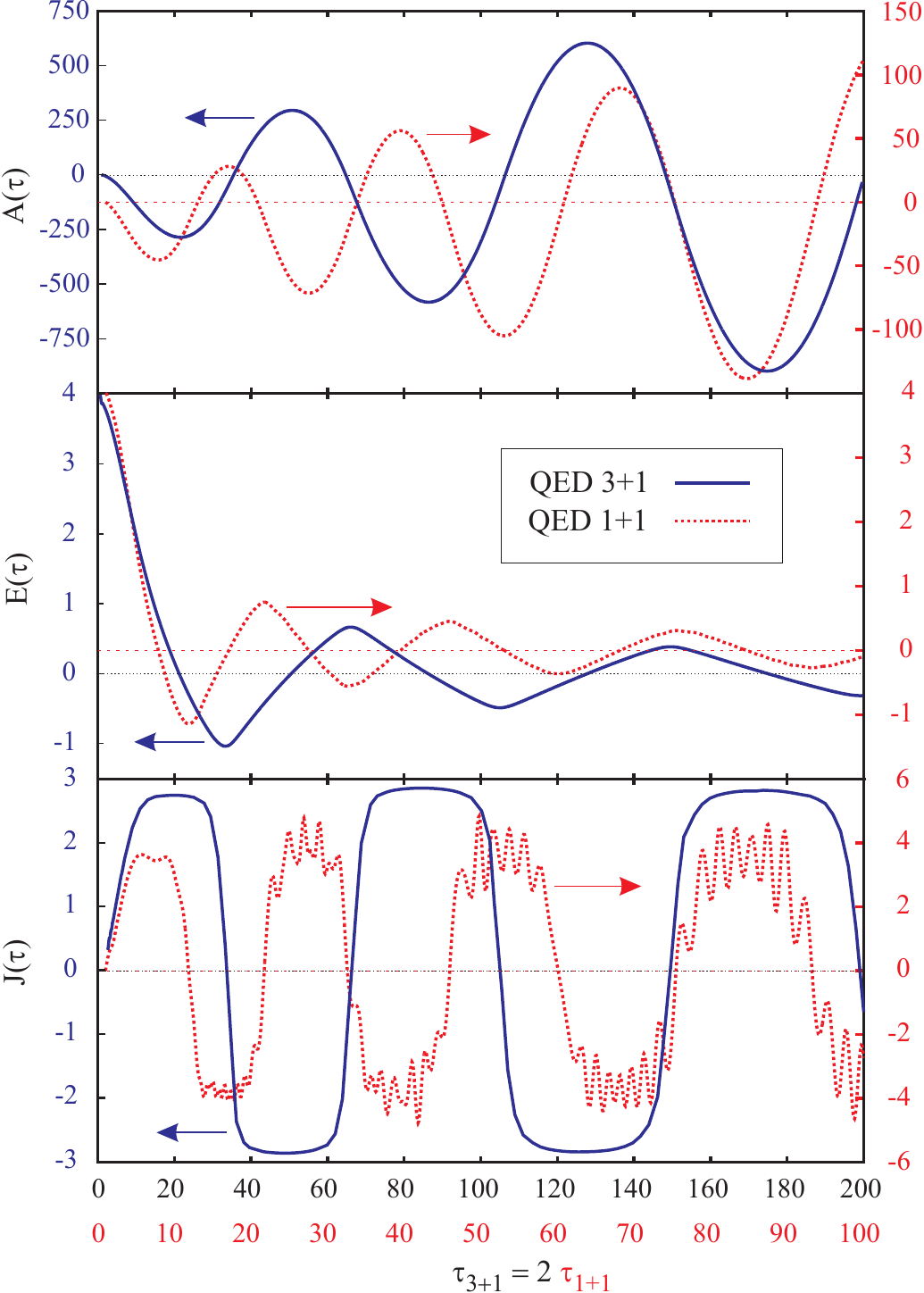}
   \caption{\label{fig4}(Color online)
   Proper-time evolution of the electromagnetic fields and current for
   the case of (1+1)- and (3+1)-dimensional QED, respectively.
   Here we choose $m=1$, $A(\tau_0)=0$ and $E(\tau_0)=4$.}
\end{figure}

\begin{figure}[t!]
   \centering
   \includegraphics[width=0.9\columnwidth]{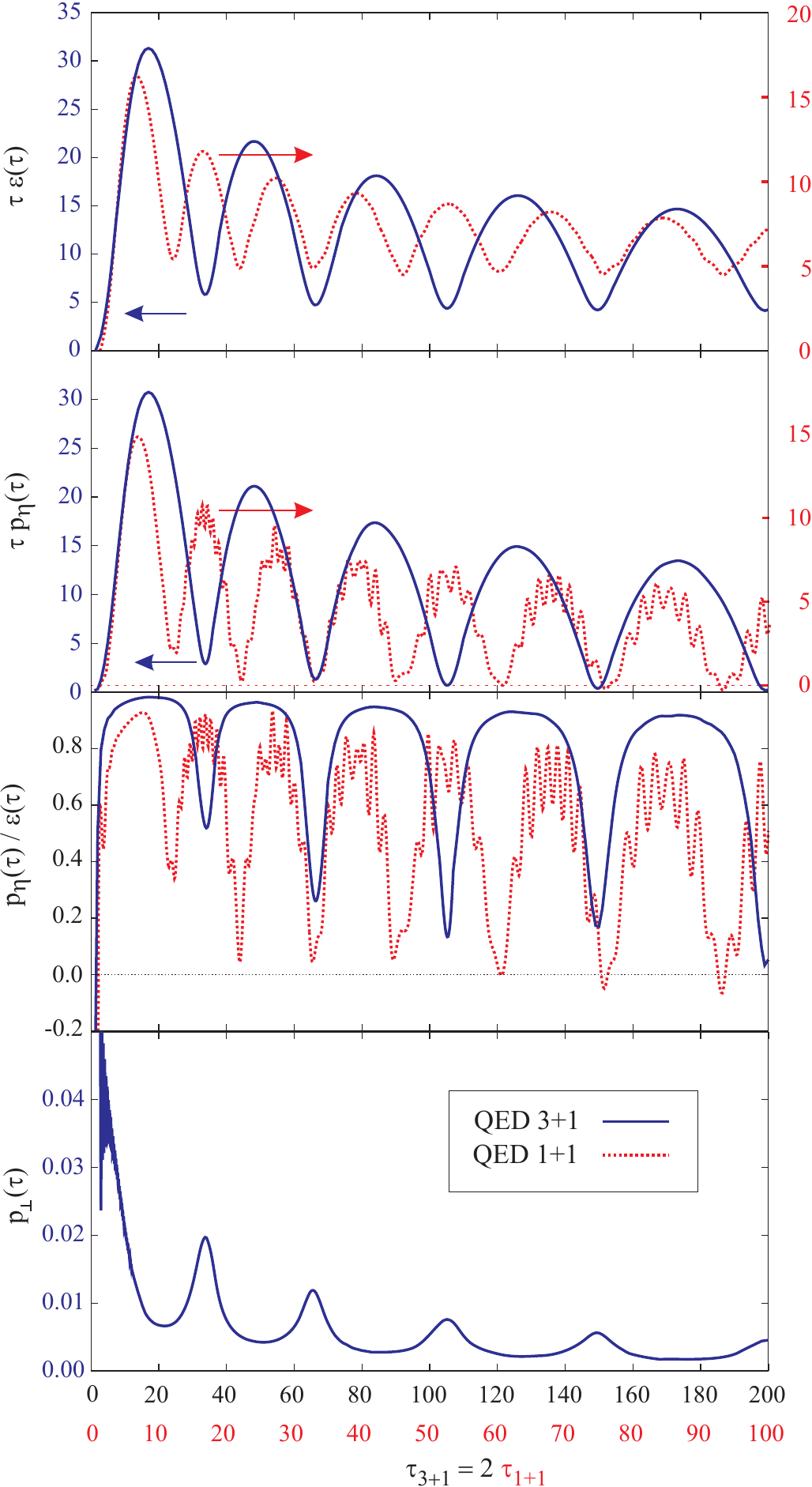}
   \caption{\label{fig5}(Color online)
   Proper-time evolution of the matter components of the renormalized
   energy-momentum tensor for the case of (1+1)- and (3+1)-dimensional QED, respectively.}
\end{figure}

\begin{figure}[t!]
   \centering
   \includegraphics[width=0.9\columnwidth]{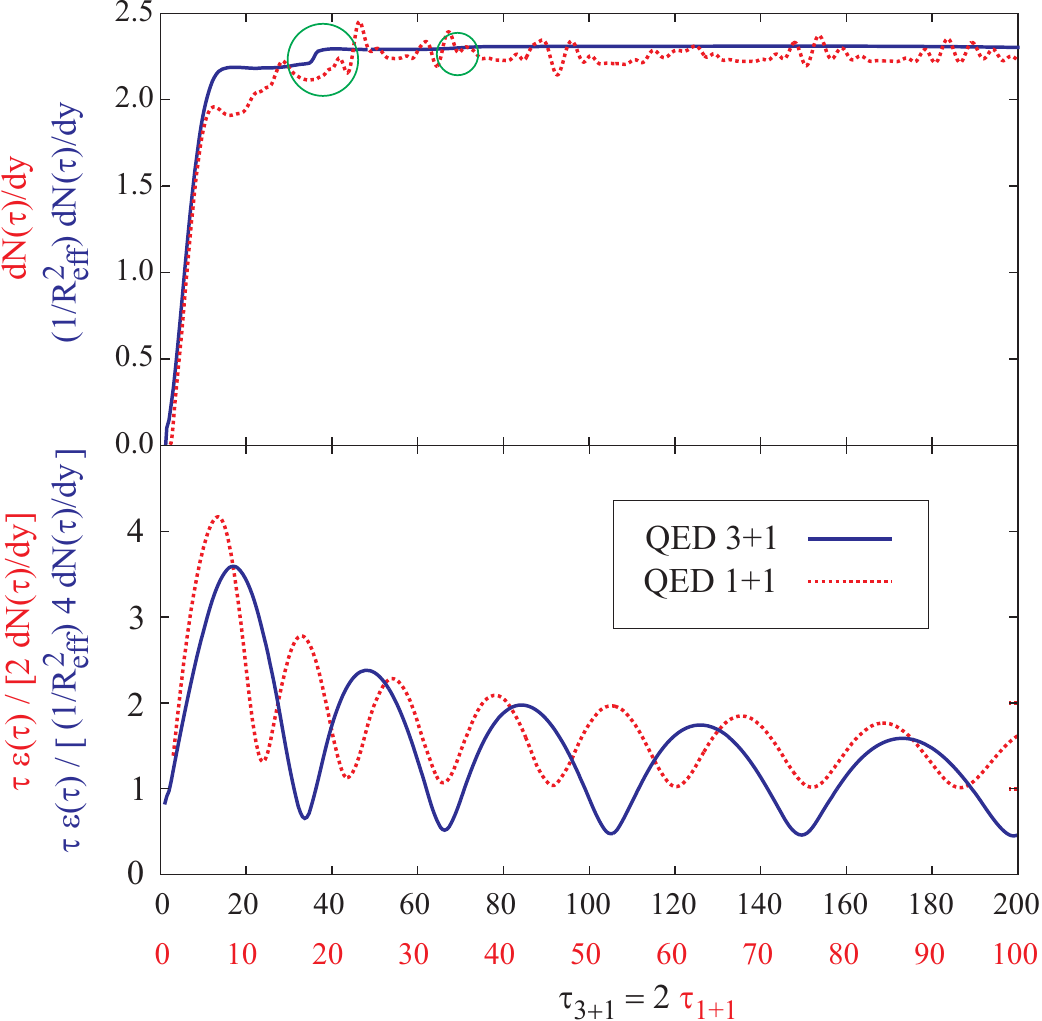}
   \caption{\label{fig6}(Color online)
   Proper-time evolution of the density of pairs,
   $\mathrm{d}N/\mathrm{d}y$,
   and proper time evolution of the ratio
   $\tau \varepsilon(\tau) / [ \mathrm{d}N/\mathrm{d}y ]$
   for the case of (1+1)- and (3+1)-dimensional QED, respectively.
   The circles in the upper panel denote step increases in the particle density corresponding
   to the second and third gradients of the current.
   }
\end{figure}

%
%

\section{Quasiparticle phase space distribution functions}
\label{s:quasi}

For the problem at hand, particle production from classical
electric fields, it is possible to introduce an interpolating
number density via a Bogoliubov transformation that is an
adiabatic invariant. This was done
previously~\cite{r:KESCMprl91,r:Kluger:1992fk,r:Cooper:1993uq,Mihaila:2008dp}).
However, when
we consider the fully interacting case with quantum gauge
fields then one needs to resort to an ``effective''
quasiparticle distribution function that allows one to
reproduce the expectation value of the current and the
energy-momentum tensor. That is, we want to determine an
effective distribution function $f(x,k)$ in analogy with
relativistic kinetic theory (see for example
Refs.~\cite{r:CFSprd75,Hu}) such that
\begin{equation}
   \Expect{ J^{\mu} }
   =  e  \int \rD k \, k^{\mu} \, f(x,k) \>,
\end{equation}
and
\begin{equation}
   \Expect{ T^{\mu \nu} }
   =
   \int \rD k \, k^{\mu} k^{\nu} \, f(x,k) \>,
\end{equation}
where
\begin{equation}
   \rD k
   =
   2 r \, \theta(k^0) \delta(k^2 - M^2) \,
   \frac{d^4 k}{(2 \pi)^3 \sqrt{-g} } \>.
\end{equation}
Here, $r$ is a degeneracy factor which counts the number of
species.  For our case of quark and anti-quark pairs with spin
one-half, we have $r=4$.
Hence, the renormalized comoving energy density,
$\mathcal{\bar E} = T_{00}$, is given by
\begin{equation}
   \mathcal{\bar E}
   =
   \frac{4}{\tau}
   \int_{0}^{+\infty} \! \frac{k_{\perp} \, \rd k_{\perp}}{2\pi} \!\!
   \int_{-\infty}^{+\infty} \! \frac{\rd k_{\eta}}{2\pi} \
   \omega_{ k_{\perp}, k_{\eta} }(\tau) \,
   f(\tau,k_{\perp},k_{\eta}) \>,
\end{equation}
and will be identified with the renormalized field theory result:
\begin{equation}
   \mathcal{E}
   =
   -
   \biggl [ \,
      \int \>
      \bk_{\kperp,h}^{\phantom{(}}(\peta)
      \cdot
      \bP^{(+)}_{\kperp,h}(\peta,\tau)
      +
      \frac{E^2}{2} \,
   \biggr ]_{\text{ren}} \>,
\end{equation}
where we have subtracted the divergences coming from the
cosmological term and the charge renormalization. Note that
when the single-particle distribution becomes independent of
proper time, $\tau$, one can easily derive the conservation of energy equation
in terms of the energy density and longitudinal pressure:
Consider the identity
\begin{equation*}
   \frac{4}{\tau} \!
   \int_{0}^{+\infty} \! \frac{k_{\perp} \, \rd k_{\perp}}{2\pi} \!\!
   \int_{-\infty}^{+\infty} \! \frac{\rd k_{\eta}}{2\pi} \,
   \omega_{ k_{\perp}, k_{\eta} }(\tau) \,
   \frac{\partial f(\tau,k_{\perp},k_{\eta}) }{\partial \tau}
   =
   0 \>,
\end{equation*}
then integrate by parts to obtain
\begin{equation}
   \frac{ \partial \mathcal{\bar E} }{\partial \tau}
   +
   \frac { \mathcal{\bar E} + \mathcal{\bar P}_{\eta} }{\tau}
   =
   0 \>,
\end{equation}
where the longitudinal pressure is introduced as
\begin{equation}
   \mathcal{\bar P}_{\eta}
   =
   \frac{4}{\tau} \!
   \int_{0}^{+\infty} \! \frac{k_{\perp} \, \rd k_{\perp}}{2\pi} \!\!
   \int_{-\infty}^{+\infty} \! \frac{\rd k_{\eta}}{2\pi} \,
   \frac{ ( \, k_{\eta} / \tau \, )^2 }
        { \omega_{ k_{\perp}, k_{\eta} }(\tau) } \,
   f(\tau,k_{\perp},k_{\eta}) \>.
\end{equation}

The quasiparticle phase-space distribution of pairs
of particles and antiparticles \textcolor{black}{with a specific spin}
in light-cone variables is introduced as
\begin{equation}\label{interplc}
   \textcolor{black}{
   \frac{ \rd^6 N }
        { \rd^2 x_{\perp} \rd \eta \
          \rd^2 k_{\perp} \rd k_{\eta} }
   =
   \frac{f(\tau, k_{\perp}, k_{\eta} )}{(2 \pi)^3}
   \>,
   }
\end{equation}
such that the pair density is obtained as
\begin{equation}
   \frac{ \rd^3 N }
        { \rd^2 x_{\perp} \rd \eta }
   =
   \int_{0}^{+\infty} \!
   \frac{\kperp \, \rd \kperp}{ 2\pi } \,
   \int_{-\infty}^{+\infty} \! \frac{\rd k_\eta}{2\pi} \
   f(\tau, k_{\perp}, k_{\eta} )
   \>.
\end{equation}
Here, we have $\rd^2 x_{\perp} = \rho \, \rd \rho \, \rd
\theta$ and $\rd^2 k_{\perp} = k_{\perp} \rd k_{\perp} \, \rd
\phi$.
For completeness, we note that in (1+1) dimensions the pair density reads
\begin{equation}
   \frac{ \rd N }
        { \rd \eta }
   =
   \int_{-\infty}^{+\infty} \! \frac{\rd k_\eta}{2\pi} \
   f(\tau, k_{\eta} )
   \label{dndy_1d}
   \>.
\end{equation}

When the pair distribution becomes independent of the proper
time $\tau$ then we are in the ``out-regime'' and can stop our
calculation as far as determining the particle spectra. We need
to relate this quantity to the center-of-mass distribution of
electrons and positrons produced by the strong electric field.
We introduce the free-particle rapidity, 
$y = \frac{1}{2} \ln[ (E+k_z)/(E-k_z)]$,
and ``transverse'' mass, 
$M_{\perp} = \sqrt{k_{\perp}^2 + m^2}$,
by relating them to the cartesian coordinate four-momentum in the center-of-mass
system, $\hat{k}^a = (E,{\bf k})$, by the relation
\begin{align*}
   \hat{k}^a
   &=
   \Set{ M_{\perp} \cosh y, {\bf k}_\perp, M_{\perp} \sinh y }
\end{align*}
The boost that takes one from the center-of-mass coordinates to
the comoving frame, where the energy momentum tensor is diagonal,
is given by $\tanh \eta= v = z/t$, so that one can define the
``fluid'' four-velocity in the center-of-mass frame as
\begin{equation}
   u^{a} = (\cosh \eta,0,0, \sinh \eta)
   \>.
\end{equation}
It is important to relate the momenta canonical to $\eta$
and $\tau$ to the center-of-mass variables. In the out regime
we can identify these canonical momenta from the free particle
(1+1)-dimensional Lagrangian in covariant form. We show
now that
\begin{equation}
   \tau
   =
   \sqrt{ t^2 - z^2 } \>,
   \qquad
   \eta
   =
   \frac{1}{2} \ln \left( \frac{t+z}{t-z} \right) \>,
\end{equation}
have as their canonical momenta
\begin{equation}\label{boost_4trans}
   k_{\tau}
   =
   E \, t / \tau - k_z \,z / \tau \>,
   \qquad
   k_{\eta}
   =
   -E \, z + t \, k_z \>.
\end{equation}
Consider the metric $ ds^2= d\tau^2 - \tau^2 d
\eta^2$ and the free particle Lagrangian in (1+1) dimensions
\begin{equation}
   L
   =
   \frac{M}{2} \,
   g_{\mu \nu} \, \frac{\rd x^{\mu}}{\rd s} \, \frac{\rd x^{\nu}}{\rd s}
   \>.
\end{equation}
Then we obtain
\begin{align}
   k_{\tau}
   &=
   M \, \frac{\rd \tau}{\rd s}
   =
   M \,
   \biggl [ \,
      \biggl ( \frac{\partial \tau}{\partial t} \biggr )_z \,
      \frac{\rd t}{\rd s}
      +
      \biggl ( \frac{\partial \tau}{\partial z} \biggr )_t \,
      \frac{\rd z}{\rd s} \,
   \biggr ]
   \label{e:XXi} \\
   &=
   \frac{1}{\tau} \, ( E \, t - k_z  \, z )
   =
   \hat{k}^{a} u_{a}
   =
   M_{\perp} \cosh(\eta - y) \>,
   \notag
\end{align}
and
\begin{align}
   k_{\eta}
   &=
   M \, \frac{\rd \eta}{\rd s}
   =
   M \,
   \biggl [ \,
      \biggl ( \frac{\partial \eta}{\partial t} \biggr )_z \,
      \frac{\rd t}{\rd s}
      +
      \biggl ( \frac{\partial \eta}{\partial z} \biggr )_t \,
      \frac{\rd z}{\rd s} \,
   \biggr ]
   \label{e:XXii} \\
   &=
   - E \, z + k_z t
   =
   - \tau \, M_{\perp} \sinh(\eta-y) \>.
   \notag
\end{align}
It follows that $k_{\tau} = \hat{k}^{\mu} \, u_{\mu} = M_{\perp} \cosh(\eta-y)$
has the meaning of the energy of the particle in the comoving
frame.

The interpolating phase-space density $f$
of particles depends on $(\tau, k_{\perp},k_{\eta})$
and is found to be $\eta$-independent.
In order to obtain the center-of-mass
particle rapidity and transverse momentum distribution, we
change variables from $(\eta, k_{\eta})$ to $(z, y)$ at a fixed proper time
$\tau$, i.e.
\begin{equation}\label{e:d3Ndetadketa}
   \frac{\rd^6 N}
        {\rd^2 x_{\perp} \rd^2 k_{\perp} \rd z \, \rd y }
   =
   \frac{\rd^6 N}
        {\rd^2 x_{\perp} \rd^2 k_{\perp} \rd \eta \, \rd k_{\eta}} \,
   \biggl | \,
      \frac{ \partial( \eta, k_{\eta} ) }{ \partial( z, y ) } \,
   \biggr |_\tau
   \>.
\end{equation}
So, from \eqref{interplc}, we have
\begin{equation}\label{e:boost_J}
   \textcolor{black}{
   \frac{\rd^3 N}
        {\rd^2 k_{\perp} \rd y }
   =
   \iint \rd^2 x_{\perp} \, \rd z \,
   \biggl | \,
      \frac{ \partial( \eta, k_{\eta} ) }{ \partial( z, y ) } \,
   \biggr |_\tau \,
   \frac{f(\tau, k_{\perp}, k_{\eta} )}{(2 \pi)^3}
   \>,
   }
\end{equation}
where the Jacobian is evaluated at a fixed proper time~$\tau$,
\begin{align}
   \biggl | \,
      \frac{ \partial( \eta, k_{\eta} ) }{ \partial( z, y ) } \,
   \biggr |_\tau
   &=
   \left| \,
      \begin{matrix}
          {\partial k_{\eta}} / {\partial y}
          &
          {\partial k_{\eta}}/{\partial z}
          \\
          {\partial \eta}/{\partial y}
          &
          {\partial \eta}/{\partial z}
      \end{matrix} \,
  \right |_\tau
  =
  \left |
     \frac{ \partial k_{\eta} }{ \partial y } \,
     \frac{ \partial \eta }{ \partial z } \,
  \right |_\tau
  \label{boost_Jinv} \\
  &=
  \frac{ M_{\perp} \cosh(\eta-y) }{ \cosh \eta } \>.
  \notag
\end{align}
However, since at fixed $\tau$, we have
\begin{equation}
   \left | \,
      \frac{\partial k_{\eta}}{\partial y} \,
   \right |_\tau
   =
   \left | \,
      \frac{\partial k_{\eta}}{\partial \eta} \,
   \right|_\tau
   \>,
\end{equation}
we obtain
\begin{equation}\label{boost_Jinv2}
   \biggl | \,
      \frac{ \partial( \eta, k_{\eta} ) }{ \partial( z, y ) } \,
   \biggr |_\tau
   =
   {\partial k_{\eta}\over \partial z} \biggr |_{\tau} \>.
\end{equation}
Calling the integration over the transverse dimensions the
effective transverse size of the colliding ions $A_{\perp} = \pi R_\mathrm{eff}^2$ we
then find from \eqref{e:boost_J} that:
\begin{equation}\label{e:thisquantity}
   \textcolor{black}{
   \frac{\rd^3 N}
        {\rd^2 k_{\perp} \, \rd y }
   =
   \frac{A_{\perp}}{(2 \pi)^3}
   \int \rd k_{\eta} \,
   f(\tau, k_{\perp}, k_{\eta} )
   \equiv
   \frac{\rd^3 N}{\rd^2 k_{\perp} \, \rd \eta}
   \>.
   }
\end{equation}
The quantity in Eq.~\eqref{e:thisquantity} is independent of
$y$ which is a consequence of the assumed boost invariance.
Therefore, using the property of the Jacobean, we have proven
that the distribution of particles in particle rapidity is the
same as the distribution of particles in fluid rapidity,
verifying that in the boost-invariant regime Landau's intuition
was correct~\cite{r:Landau:1953ys}.

We now want to motivate the Cooper-Frye formula used to
calculate particle spectrum in hydrodynamical models of
particle production~\cite{r:CFSprd75}.  We have that a constant
$\tau$ surface, which is the freeze-out surface of Landau, is
parametrized as
\begin{align}
   \rd \Sigma^{a}
   &=
   A_{\perp} \, \Set{\rd z,0,0,\rd t}
   \\
   &=
   A_{\perp} \, \tau \, \rd\eta \,
   \Set{\cosh \eta ,0,0,\sinh \eta} \>.
   \notag
\end{align}
Therefore, we find
\begin{align}
   \hat{k}^{a} \, \rd \Sigma_{a}
   &=
   A_{\perp} \, M_{\perp} \, \tau \, \cosh(\eta-y) \, \rd \eta
   \\
   &=
   A_{\perp} \, | \, \rd k_{\eta} \, | \>.
   \notag
\end{align}
Thus, we can rewrite our expression for the field theory particle
spectra as
\begin{align}
   \textcolor{black}{
   \frac{\rd^3 N}{\rd^2 k_{\perp} \, \rd y }
   }
   &=
   \frac{A_{\perp}}{(2 \pi)^3}
   \int \rd k_{\eta} \,
   f(\tau, k_{\perp}, k_{\eta} )
   \\
   &=
   \int
   \hat{k}^{a} \,  \rd \Sigma_{a} \,
   f(\tau,k_{\perp},k_{\eta}) \>,
   \notag
\end{align}
where in the integration we keep $y$ and $\tau$ fixed.  Thus,
with the replacement of the thermal single-particle distribution by
the quasiparticle distribution function, we get via the coordinate
transformation to the center-of-mass frame the Cooper-Frye formula.
For completeness, we note that in (1+1) dimensions the particle spectra,
$\rd N/\rd y$, are given by the integral in Eq.~\eqref{dndy_1d}.

The boost invariant assumption leads to an energy momentum
tensor which is diagonal in the ($\tau, \rho, \theta, \eta$)
coordinate system which is thus a comoving one.  In that system
one has for the matter energy-momentum tensor
\begin{equation}
   T_{\mu\nu}
   =
   \Diag{\mathcal{E},
         \mathcal{P}_{\rho},
         \rho^2 \mathcal{P}_{\theta},
         \tau^2 \mathcal{P}_{\eta} } \>.
\end{equation}
Thus we find in this approximation that there are two separate
pressures, one in the longitudinal direction and one in the
transverse direction which is quite different from the thermal
equilibrium case. However only the longitudinal pressure enters
into the energy conservation equation:
\begin{equation}
   \frac{ \rd ( \tau \mathcal{E} ) }{\rd \tau}
   +
   \mathcal{P}_{\eta}
   =
   E \, J_{\eta} \>.
\end{equation}
It is useful to rewrite the conservation of
energy in the out regime as :
\begin{equation}
   \frac{ \rd \mathcal{E} }{\rd \tau}
   +
   \frac{ \mathcal{E} + \mathcal{P}_{\eta}}{\tau}
   =
   0 \>. \label{conservation}
\end{equation}
So to the extent that the ultra-relativistic one-dimensional
equation of state $\mathcal{E} = \mathcal{P}_{\eta}$ is true,
then one has the simple result
\begin{equation}
   \mathcal{E}
   \propto
   \tau^{-2} \>.
\end{equation}
It turns out, as our numerical results show below, that
although $p_\eta/\varepsilon \approx 1$  for part of the period
of the oscillation, during the minima $ p_\eta \rightarrow 0$
and this seems to be governing the falloff which is more like
$\varepsilon \propto 1/\tau$.

%
%

\begin{figure*}[t!]
   \centering
   \includegraphics[width=0.7\textwidth]{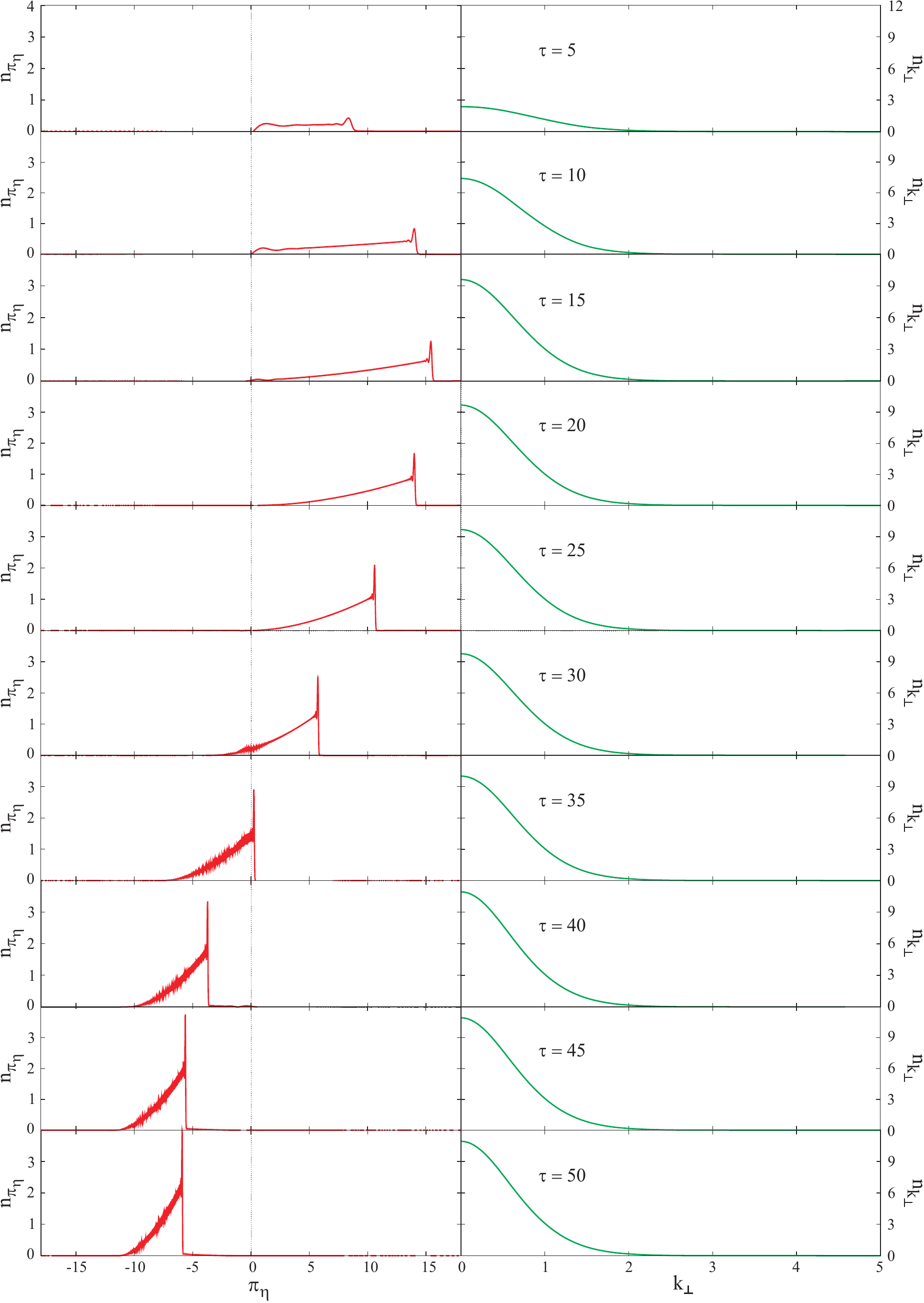}
   \caption{\label{fig7}(Color online)
   Proper-time evolution of the longitudinal momentum-dependent pair-density
   distribution, $n_{\peta}$, and the proper-time evolution of the transverse
   momentum-dependent pair-density distribution, $n_{k_\perp}$.
   (See also Ref.~\onlinecite{ref:qed3dmovies})}
\end{figure*}

\begin{figure}[t!]
   \centering
   \includegraphics[width=0.9\columnwidth]{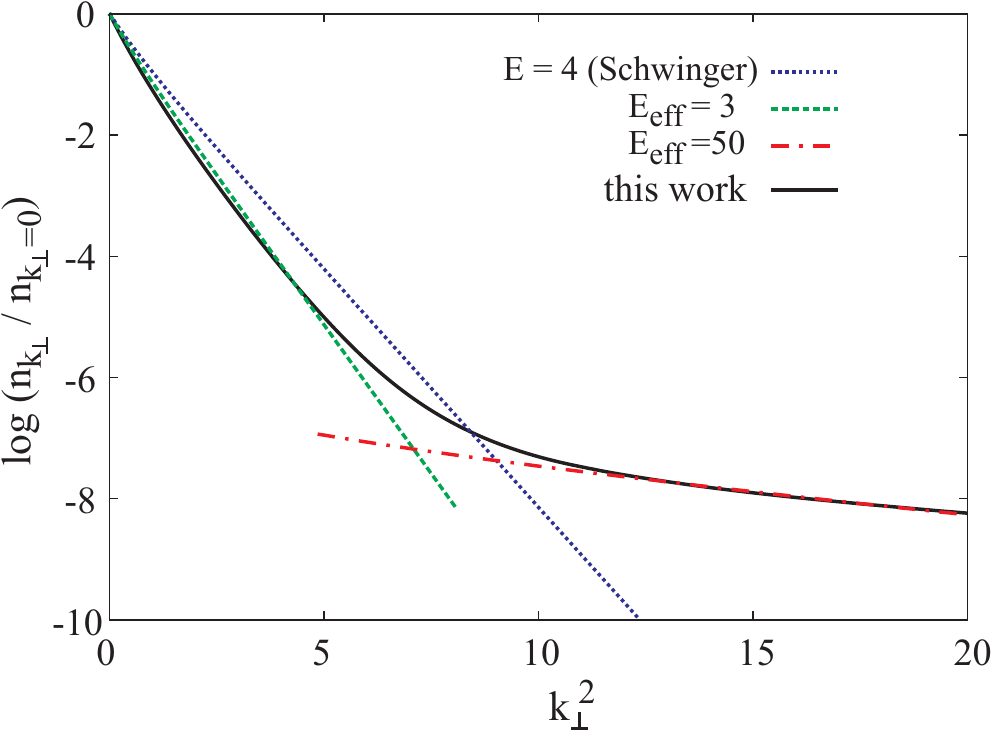}
   \caption{\label{fig8}(Color online)
   Comparison of the transverse distribution of particles according to our numerical
   simulation and the transverse distribution of particles for the constant
   electromagnetic field case with $eE = 4$ as described by Eq.~\eqref{i.e:dsf}.
   }
\end{figure}

\section{Numerical results}
\label{s:results}

Let us review the equations we intend to solve numerically. The
first of these is the polarization equation~\eqref{e:eompol}
\begin{equation}\label{nr.e:eompol}
   \partial_{\tau} \bP_{k}^{(\lambda)}(\tau)
   =
   2 \, \bk_{k}(\tau) \times \bP_{k}^{(\lambda)}(\tau) \>,
\end{equation}
and the second is the backreaction equation \eqref{e:maxwellsub}
\begin{align}
   &\partial_{\tau} E(\tau)
   =
   \frac{e}{1 + e^2 ( \delta e^2 ) }
   \sum_{p}^{\Lambda}
   \biggl [ \,
      P_{1;\kperp,h}^{(+)}(\peta,\tau)
      \label{nr.e:maxwellsub} \\ & \qquad
      -
      \frac{\peta}{\omega}
      +
      \frac{ e E(\tau) }{\tau} \,
      ( \, \kperp^2 + M^2 \, ) \,
      \Bigl ( \,
         \frac{1}{4 \, \omega^5}
         -
         \frac{5 \, \peta^2}{4 \, \omega^7} \,
      \Bigr )
   \biggr ] \>.
   \notag
\end{align}
Here we have subtracted from the integral the adiabatic
expansion of $P_{1;\kperp,h}^{(+)}(\peta,\tau)$.

In order to solve the coupled Dirac and backreaction equations,
we construct a grid in $\kperp$ and $\keta$ space as follows:
The $k_\eta$-momentum variable is discretized on a nonuniform
piece-wise momentum grid with a cutoff at $k_\eta =
\Lambda_\eta$; we find that a value of $\Lambda_k \approx 500$
is necessary to obtain numerical results insensitive with
respect to the cutoff.  A similar nonuniform grid is used to
discretize the $k_\perp$ variable, $k_\perp \in [0,\Lambda]$. A
fourth-order Runge-Kutta method is employed to solve the
coupled Dirac equation and backreaction problem.

For the purpose of calculating the subtracted values of the
current $J(\tau)$ and the components of the matter
energy-momentum tensor, we compute the momentum integrals
symmetrically with respect to the variable~$\peta$ rather than
$k_\eta$. The corresponding momentum cutoff in $\peta$-space is
chosen to be 20\% greater than $\tau_\mathrm{max} \Lambda_\eta$
to allow for possible very large values of $A(\tau)$, the
latter being unknown at the beginning of the calculation.

The momentum integrals with respect to  $\peta$ and $k_\perp$
are performed using a Chebyshev integration method with
spectral convergence~\cite{r:MM02}. Using this procedure, we
found that a grid of approximately 8000 points in the $k_\eta$
(or $\peta$) variable and 128 grid points in the $k_\perp$
variable is necessary to obtain a converged numerical result.
As such, the calculations for the backreaction problem in
(3+1)-dimensional QED require at least 100~times larger storage
and computational time then the corresponding (1+1)-dimensional
QED problem.

For illustrative purposes, we took: $m=1$, $e=1$, $\tau_0 = 1/m
= 1$, $A(\tau_0)=0$, and $E(\tau_0)=4$.  These strong-field
initial conditions have been shown to produce sufficient
fermion pairs at $\tau = \tau_0$ for plasma oscillations to
take place.  Just like in the (1+1)-dimensional case, the
conservation of the energy-momentum tensor, see
Eq.~\eqref{em.e:ConsI}, serves as a numerical test: for the
results of simulations reported here, the renormalized
energy-momentum tensor is conserved within machine precision.

In order to keep the size of the simulation to a minimum, we
chose the initial conditions corresponding to the
\emph{one-field} scenario introduced first in
Ref.~\cite{Mihaila:2008dp} and summarized in
Sec.~\ref{ss:initial}. In Figs.~\ref{fig1} we illustrate the
convergence of our results with respect to the choice of the
cutoff, $\Lambda$. For completeness, we depict the proper-time
evolution of the fields, $A(\tau)$ and $E(\tau)$, current,
$J(\tau)$, energy, $\epsilon(\tau)$ and transverse pressure,
$p_\eta(\tau)$, for cutoff values between 2 and~6. We conclude
that for $\Lambda=5$ the results are insensitive to the cutoff
$\Lambda$, within numerical accuracy.

It is important to note that our choice of initial conditions
results in a time evolution that is not consistent with the
adiabatic expansion for early values of the proper time. In
Fig.~\ref{fig2} we depict the proper-time evolution of the
fields, $A(\tau)$ and $E(\tau)$, and current, $J(\tau)$, at
early times for several values of the cutoff~$\Lambda$. We note
that while the proper-time dynamics converges for $\tau > 5$,
for earlier times the proper-time evolution depends on the
choice of the cutoff~$\Lambda$. However, for larger proper-time
values, the nonadiabatic components of the current dissipate
and the adiabatic-expansion-based subtraction becomes exact.
This behavior is illustrated numerically in Fig.~\ref{fig3},
where we depict the $k_\perp$-projected distribution of the
current for $\tau$ values of 2, 3 and~9. We notice that the
nonadiabatic oscillations of the current present at early
proper times dampen out and disappear at later proper times.
Therefore, the early proper-time evolution will be disregarded
as ``unphysical.'' This is a small price to pay in order to
keep the storage and time requirements of our simulation to a
minimum.

In Fig.~\ref{fig4}, we compare the proper-time evolution of the
electromagnetic field, $A(\tau)$, electric field, $E(\tau)$,
and current, $J(\tau)$, for (1+1)- and (3+1)-dimensional QED,
respectively. Similarly, in Fig.~\ref{fig5} we depict the
proper-time evolution of the matter components of the
energy-momentum tensor. We note that in (1+1)~dimensions the
fields evolve much faster than in (3+1)~dimensions. Also, in
(3+1)-dimensional QED the energy and longitudinal-pressure
densities are very very close in magnitude, which in turn
results in a small transverse pressure, $p_\perp(\tau)$.
Qualitatively, we also note that the modulation observed in the
proper-time evolution of the current and longitudinal pressure
in (1+1)-dimensional QED are not present any longer in
(3+1)~dimensions. Also by inspecting the two upper panels in
Fig.~\ref{fig5}, we notice that in $3+1$ dimensions $p_\eta
\approx \varepsilon$. However, the ratio $p_\eta/\varepsilon$
becomes close to zero near the minimum of the oscillation and
this in turn leads to $\tau \varepsilon$ to be almost constant
instead of going as $1/\tau$ using the arguments coming from
the energy conservation equation Eq.~\eqref{conservation}.

Finally, the proper-time evolution of the density of pairs,
$\mathrm{d}N/\mathrm{d}y$, are
depicted in Fig.~\ref{fig6}. Particles are being created
corresponding to the current gradients, with the major
contribution corresponding to the initial current gradient, and
subsequent smaller step increases before the particle density
saturates. At late values of the proper time, the ratios
$\tau \varepsilon(\tau) / [ \mathrm{d}N/\mathrm{d}y ]$
are seen to approach a constant consistent
with the hydrodynamical picture, which relates the energy  in a
bin of rapidity divided by the energy of a single particle with
that rapidity  with the number of particles in a bin of
rapidity  as explained in Ref.~\onlinecite{r:Cooper:1993uq}. In
the real problem we expect that interactions between the
fermions will eliminate the oscillations observed here.

The proper-time evolution of the momentum-dependent
longitudinal pair-density distribution, $n_{\peta}$,
defined as
\begin{equation}
   \textcolor{black}{
   n_{\peta}
   =
   \frac{2\pi}
        {A_{\perp}} \
   \frac{\rd^2 N}
        {\rd k_{\eta} \, \rd y }
   =
   \int_{0}^{+\infty} \!
   \frac{\kperp \, \rd \kperp}{ 2\pi } \,
   f(\tau, k_{\perp}, k_{\eta} )
   \>,
   }
\end{equation}
and the transverse pair-density distribution,
$n_{k_\perp}$, defined as
\begin{equation}
   \textcolor{black}{
   n_{k_\perp} =
   \frac{(2\pi)^2}
        {A_{\perp}} \
   \frac{\rd^3 N}
        {\rd^2 k_{\perp} \, \rd y }
   =
   \int_{-\infty}^{+\infty} \! \frac{\rd k_\eta}{2\pi} \,
   f(\tau, k_{\perp}, k_{\eta} )
   \>,
   }
\end{equation}
in (3+1)-dimensional QED are shown in Fig.~\ref{fig7}. (See
also Ref.~\onlinecite{ref:qed3dmovies}.) We note that the
centroid of the particle-density distribution, $n_{\peta}$,
oscillates between positive and negative values of $\peta$,
similar to the (1+1)-dimensional QED case.

In  Fig.~\ref{fig8} we compare the transverse momentum
distribution given by the constant field exact solution  Eq.
(\ref{i.e:dsf}) with the results of our numerical solution for
$eE=4$.  Part of the results are expected, that is at small
transverse momenta the distribution of particles is similar to
the static case but with a smaller effective field since the
field is decreasing during the first phase of particle
production. What is unexpected is that, in the problem with
backreaction,  there is a new tail in the transverse momentum
distribution which falls exponentially with an effective $|eE|
= 50$.   This is a totally surprising result whose origin we do
not yet have a simple explanation for.

%
%

\section{Conclusions}
\label{s:conclusions}

We have for the first time calculated the transverse
distribution of jets produced by an initial strong electric
field including the effects of backreaction. We have compared
the results of our (3+1)-dimensional calculations (for
``hydrodynamic'' quantities  as well as for  the proper time
evolution of the electric field and current) with their
(1+1)-dimensional counterparts. We find that the electric field
degrades much quicker in (3+1) dimensions than in (1+1)
dimensions. Also secondary oscillations in the current and in
the longitudinal pressure, present in (1+1) dimensions seem to
be absent in (3+1) dimensions suggesting that the extra degrees
of freedom perform some smoothing. We now have the first
numerical results for the transverse momentum distribution
function of fermion pairs which we can compare with the exact
results for the constant field problem.  We find that unlike
the constant field case, the distribution is bimodal. At modest
$k_\perp^2 \leq 5 m^2$  the transverse distribution is similar
to the constant field case with a reduced (75\%) effective $eE$
for $eE_0=4$. For larger transverse momentum $k_\perp^2 \geq
10 m^2$  the transverse distribution function has a tail
described by an effective $eE$ which is of the order of $50$.
This is a totally new feature that is as of yet not understood
simply. In a related paper~\cite{bv} we will also consider a
transport approach to the (3+1)-dimensional problem and show
that such a semiclassical picture works better in (3+1) than in
(1+1) dimensions.

%
%

\begin{acknowledgments}
This work was performed in part under the auspices of the
United States Department of Energy. The authors would like to
thank the Santa Fe Institute for its hospitality during the
completion of this work.
\end{acknowledgments}

%
\appendix
%
\section{Notation}
\label{s:notation}

In this appendix, we list our notation and conventions used
throughout this paper. We use a boldface $\bk$ to designate
the complete set of mode variables, $k$ and $k_m$ subsets of
the full set, and $k_{\phi}$ the cylindrical coordinate set.
In addition, the set $p$ substitutes the $\peta$ kinetic
momentum for $\keta$.  These sets are given by
\begin{subequations}\label{nt.e:kdefs}
\begin{align}
   \bk
   &\defby
   \Set{ \kperp, h, \keta, m } \>,
   \label{nt.e:bkmdef} \\
   k
   &\defby
   \Set{ \kperp, h, \keta } \>,
   \label{nt.e:kdef} \\
   p
   &\defby
   \Set{ \kperp, h, \peta } \>,
   \label{nt.e:pdef} \\
   k_m
   &\defby
   \Set{ \kperp, m } \>,
   \\
   \kphi
   &\defby
   \Set{ k_x, k_y, \keta }
   =
   \Set{ \kperp, \phi, \keta } \>.
   \label{nt.e:bkphidef}
\end{align}
\end{subequations}
Sums over these quantities indicate the following integrals and sums
\begin{subequations}\label{nt.e:ksumdefs}
\begin{align}
   \sum_{\bk}
   &\defby
   \int_{0}^{+\infty}
      \frac{\kperp \, \rd \kperp}{ 2\pi } \!\!
   \int_{-\infty}^{+\infty}
      \frac{\rd \keta}{2\pi}
   \sum_{m =-\infty}^{+\infty} \>
   \sum_{h = \pm 1} \>,
   \label{nt.e:bksumdef} \\
   \sum_{k}
   &\defby
   \int_{0}^{+\infty} \!
   \frac{\kperp \, \rd \kperp}{ 2\pi } \,
   \int_{-\infty}^{+\infty} \! \frac{\rd \keta}{2\pi} \,
   \sum_{h=\pm 1} \>,
   \label{nt.e:ksumdef} \\
   \sum_{p}
   &\defby
   \int_{0}^{+\infty} \!
   \frac{\kperp \, \rd \kperp}{ 2\pi } \,
   \int_{-\infty}^{+\infty} \! \frac{\rd \peta}{2\pi} \,
   \sum_{h=\pm 1} \>,
   \label{nt.e:psumdef} \\
   \sum_{\kphi}
   &\defby
   \int_{-\infty}^{+\infty} \frac{ \rd k_x}{ 2\pi }
   \int_{-\infty}^{+\infty} \frac{ \rd k_y}{ 2\pi }
   \int_{-\infty}^{+\infty} \frac{ \rd \keta}{ 2\pi }
   \\
   &=
   \int_{0}^{\infty} \frac{ \kperp \, \rd\kperp }{2\pi}
   \int_{0}^{2\pi} \frac{\rd\phi}{2\pi}
   \int_{-\infty}^{+\infty} \frac{ \rd \keta}{ 2\pi } \>.
   \label{nt.e:kphisumdef}
\end{align}
\end{subequations}
We also use the following notation for $\delta$-functions
\begin{equation}\label{nt.e:kdeltadef}
   \delta_{\bk,\bk'}^{\phantom(}
   \defby
   ( 2\pi )^2 \,
   \delta_{h,h'} \,
   \delta_{m,m'} \,
   \frac{ \delta(\kperp - \kperp') }{ \sqrt{ \kperp \, \kperp' } } \,
   \delta(\keta - \keta') \>.
\end{equation}
In a similar way, we put $\mathbf{x} \defby \Set{ \rho, \theta, \eta }$.  The sum over $\mathbf{x}$ means
\begin{equation}\label{nt.e:sumxdef}
   \sum_{\mathbf{x}}
   \defby
   \int_{0}^{\infty} \!\! \frac{\rho \, \rd\rho}{2\pi}
   \int_{0}^{2\pi} \!\! \frac{\rd\theta}{2\pi} \,
   \int_{-\infty}^{+\infty} \! \frac{\rd\eta}{2\pi} \>,
\end{equation}
and $\delta_{\mathbf{x},\mathbf{x}'}$ means
\begin{equation}\label{nt.e:xdeltadef}
   \delta_{\mathbf{x},\mathbf{x}'}
   \defby
   \frac{ \delta(\rho - \rho' ) }{ \sqrt{\rho \rho'} } \,
   \delta(\theta - \theta') \,
   \delta( \eta - \eta' )  \>.
\end{equation}

%
%
\section{Transverse helicity eigenvectors}
\label{s:transhelicity}

In this section we derive transverse helicity eigenvectors and
show how they can be used to expand solutions of the Dirac
equation in boost-invariant coordinates.  The Hermitian
two-component transverse helicity operator $H_{\perp}$ is
defined in momentum space by
\begin{equation}\label{h.e:Htwotransdef}
   H_{\perp}
   =
   \frac{1}{\kperp} \,
   ( \, \mathbf{k} \times \bsigma  \, )
   \cdot \hat{\mathbf{e}}_z
   =
   \begin{pmatrix}
      0 & -i e^{-i\phi} \\
      +i e^{+i\phi} & 0
   \end{pmatrix} \>,
\end{equation}
where $k_x = \kperp \cos \phi$ and $k_y = \kperp \sin \phi$.
We write the eigenvalue equation for this operator as:
\begin{equation}\label{h.e:Heigenequ}
   H_{\perp} \, \chi_{\phi,h}
   =
   h \, \chi_{\phi,h} \>,
\end{equation}
with eigenvalues $h = \pm 1$ and orthogonal eigenvectors:
\begin{equation}\label{h.e:chivectors}
   \chi_{\phi,h}
   =
   \frac{1}{\sqrt{2}} \,
   \begin{pmatrix}
      1
      \\
      i h \, e^{i\phi}
   \end{pmatrix} \>.
\end{equation}
We note that $\sigma_z \, \chi_{\phi,h} = \chi_{\phi,-h}$, and that:
\begin{equation}\label{h.e:sxsyonchi}
   \bigl ( \,
      \sigma_x \, k_x
      +
      \sigma_y \, k_y \,
   \bigr ) \, \chi_{\phi,h}
   =
   i h \kperp \, \chi_{\phi,-h} \>.
\end{equation}

For the coordinate system $x^{\mu} = ( \, \tau, x, y, \eta \, )$, Dirac's equation is:
\begin{equation}\label{h.e:DiracI}
   \bigl \{ \,
      \tilde{\gamma}^{\mu}(\tau,\eta) \,
      [ \, i \partial_{\mu} - g \, A_{\mu}(\tau) \, ]
      -
      M \,
   \bigr \} \, \psi(\tau, x, y, \eta)
   =
   0 \>,
\end{equation}
where now:
\begin{equation*}
\begin{split}
   \tilde{\gamma}^{\tau}(\eta)
   &=
      \cosh\eta \, \gamma^0
      -
      \sinh\eta \, \gamma^3 \,
   \>,
   \\
   \tilde{\gamma}^{\eta}(\tau,\eta)
   &=
   \bigl ( \,
      -
      \sinh\eta \, \gamma^0
      +
      \cosh\eta \, \gamma^3 \,
   \bigr ) / \tau \>.
\end{split}
\end{equation*}
with $\tilde{\gamma}^{x} = \gamma^1$ and $\tilde{\gamma}^{y} =
\gamma^2$.  We next boost to a coordinate system where $\eta =
0$ by setting:
\begin{equation}\label{h.e:psitophi}
   \psi(\tau, x, y, \eta)
   =
   S(\eta) \,
   \phi'(\tau, x, y, \eta) / \sqrt{\tau} \>.
\end{equation}
then \eqref{h.e:DiracI} becomes
\begin{multline}\label{h.e:DiracIII}
   \bigl \{ \,
   i \, \gamma^0 \,
   \partial_{\tau}
   +
   i \,  \gamma^1 \, \partial_x
   +
   i \, \gamma^2 \, \partial_y
   \\
   +
   \gamma^3 \,
   \bigl [ \,
      i \, \partial_{\eta}
      +
      g \, A(\tau) \,
   \bigr ] / \tau
   -
   M \,
   \bigr \} \, \phi'(\tau,x,y,\eta)
   =
   0 \>,
\end{multline}
which is what we want to solve.  So let us first introduce the Fourier transform:
\begin{equation}
   \phi'(\tau, x, y, \eta)
   =
   \phi_{\bk}'(\tau) \,
   e^{ i ( k_x x + k_y y + \keta \eta ) } \>,
\end{equation}
where $\bk = (\keta,k_x,k_y)$.  Then \eqref{h.e:DiracIII} becomes:
\begin{equation}\label{h.e:DiracIV}
   \bigl [ \,
   i \, \gamma^0 \,
   \partial_{\tau}
   -
   \gamma^1 \, k_x
   -
   \gamma^2 \, k_y
   -
   \gamma^3 \, \peta(\tau)
   -
   M \,
   \bigr ] \,
   \phi'_{k}(\tau)
   =
   0 \>,
\end{equation}
where $\peta(\tau) = [ \, \keta - e A(\tau) \, ] / \tau$ is the
kinetic momentum.  Using Eq.~\eqref{h.e:sxsyonchi}, we see that
\eqref{h.e:DiracIV} is separable if we put
\begin{equation}\label{h.e:phimatdef}
   \phi'_{k}(\tau)
   =
   \begin{pmatrix}
      \phi_{(+);k}(\tau) \,
      \chi_{\phi,+h}
      \\
      \phi_{(-);k}(\tau) \,
      \chi_{\phi,-h}
   \end{pmatrix} \>,
\end{equation}
where $\phi_{(\pm);k}(\tau)$ now satisfy the two-component equation
\begin{multline*}
   i \partial_{\tau}
   \begin{pmatrix}
      \phi_{(+);k}(\tau)
      \\
      \phi_{(-);k}(\tau)
   \end{pmatrix}
   \\
   =
   \begin{pmatrix}
      + M
      &
      \peta(\tau) - i h \kperp
      \\
      \peta(\tau) + i h \kperp
      &
      - M
   \end{pmatrix}
   \begin{pmatrix}
      \phi_{(+);k}(\tau)
      \\
      \phi_{(-);k}(\tau)
   \end{pmatrix} \>,
\end{multline*}
which agrees with Eq.~\eqref{e:phipmeom}.  Near $\tau =
\tau_0$, there are positive and negative energy solutions to
these equations which we label by $\lambda = \pm 1$.  So the
fermi field $\hat{\phi}(\tau,x,y,\eta)$ can be expanded as
\begin{multline}\label{h.e:psiI}
   \hat{\phi}'(\tau,x,y,\eta)
   =
   \sum_{\kphi} \sum_{h=\pm 1} \sum_{\lambda=\pm 1}
   \hat{A}_{\kphi,h}^{(\lambda)} \,
   \\ \times
   \begin{pmatrix}
      \phi_{(+);k}^{(\lambda)}(\tau) \,
      \chi_{\phi,+h}
      \\
      \phi_{(-);k}^{(\lambda)}(\tau) \,
      \chi_{\phi,-h}
   \end{pmatrix}
   e^{ i ( \, \keta \eta + k_x \, x + k_y \, y \, ) } \>.
\end{multline}
where $\hat{A}_{\kphi,h}^{(\lambda)}$ are the creation and
annihilation operators for the state described by
$(\kphi,h,\lambda)$. Now let us introduce cylindrical
coordinates, $x = \rho \, \cos \theta$ and $y = \rho \, \sin
\theta$, so that
\begin{equation}\label{h.e:xykxky}
\begin{split}
   k_x \, x + k_y \, y
   &=
   \kperp \rho \,
   \cos ( \theta - \phi ) \>.
\end{split}
\end{equation}
Now the generating function for Bessel functions is given by
\begin{equation}\label{h.e:bessgenIx}
   \Exp{ z ( t - 1/t )/2 }
   =
   \sum_{m=-\infty}^{+\infty}
   t^m \, J_m(z) \>.
\end{equation}
If we put $t = i e^{i (\theta - \phi) }$ and $z=\kperp \, \rho$, this becomes
\begin{equation*}
\begin{split}
   \Exp{ i \kperp \rho \, \cos ( \theta - \phi ) }
   =
   \sum_{m=-\infty}^{+\infty}
   i^{m} \,
   e^{i m \, (\theta - \phi)} \,
   J_{m}(\kperp\rho) \>.
\end{split}
\end{equation*}
Using these results in Eq.~\eqref{h.e:psiI}, we find in cylindrical coordinates the expansion,
\begin{multline}\label{h.e:psiII}
   \hat{\phi}'(\tau,\rho,\theta,\eta)
   =
   \sum_{\kphi}
   \sum_{m=-\infty}^{+\infty}
   \sum_{h=\pm 1} \sum_{\lambda=\pm 1}
   \hat{A}_{\kphi,h}^{(\lambda)} \,
   e^{ i \keta \eta }
   \\ \times
   \begin{pmatrix}
      \phi_{(+);k}^{(\lambda)}(\tau) \,
      \chi_{\phi,+h}
      \\
      \phi_{(-);k}^{(\lambda)}(\tau) \,
      \chi_{\phi,-h}
   \end{pmatrix}
   i^{m} \,
   e^{i m \, (\theta - \phi)} \,
   J_{m}(\kperp\rho) \>.
\end{multline}
Now let us define the Fourier transform pair:
\begin{subequations}\label{h.e:ftA}
\begin{align}
   i^m
   \hat{A}_{\keta,\kperp,m,h}^{(\lambda)}
   &=
   \int_{0}^{2\pi} \frac{ \rd \phi}{ 2\pi } \,
   \hat{A}_{\keta,\kperp,\phi,h}^{(\lambda)} \,
   e^{-i m \phi} \>,
   \label{h.e:ftAa} \\
   \hat{A}_{\keta,\kperp,\phi,h}^{(\lambda)}
   &=
   \sum_{m=-\infty}^{+\infty}
   i^m
   \hat{A}_{\keta,\kperp,m,h}^{(\lambda)} \,
   e^{i m \phi} \>.
   \label{h.e:ftAb}
\end{align}
\end{subequations}
So using \eqref{h.e:ftAa}, and putting $m \rightarrow m + 1$ in
the second and fourth components, Eq.~\eqref{h.e:psiII} becomes
\begin{multline}\label{h.e:psiIII}
   \hat{\phi}'(\tau,\rho,\theta,\eta)
   =
   \sum_{\bk} \sum_{\lambda=\pm 1}
   \hat{A}_{\bk}^{(\lambda)} \;
   \\ \times
   e^{ i \keta \eta }
   \begin{pmatrix}
      \phi_{(+);k}^{(\lambda)}(\tau) \,
      \chi_{k_m,+h}'(\rho,\theta)
      \\[3pt]
      \phi_{(-);k}^{(\lambda)}(\tau) \,
      \chi_{k_m,-h}'(\rho,\theta)
   \end{pmatrix} \>.
\end{multline}
where now
\begin{equation}\label{h.e:chirhothetadef}
   \chi_{k_m,h}'(\rho,\theta)
   =
   \frac{1}{\sqrt{2}}
   \begin{pmatrix}
      e^{i m \theta} \, J_{m}(\kperp\rho)
      \\
      - h \, e^{i (m+1) \theta} \, J_{m+1}(\kperp\rho)
   \end{pmatrix} \>.
\end{equation}
Finally, we boost to a coordinate system where $\theta=0$ by
multiplying $\hat{\phi}'(\tau,\rho,\theta,\eta)$ by
$S_{\rho}^{-1}(\theta)$, which can be written as
\begin{equation}\label{h.e:Srho}
\begin{split}
   S_{\rho}^{-1}(\theta)
   &=
   \cos( \theta/2 ) - \gamma^1 \gamma^2 \, \sin( \theta /2 )
   \\
   &=
   \begin{pmatrix}
      e^{+i\theta/2} & 0 & 0 & 0 \\
      0 & e^{-i\theta/2} & 0 & 0 \\
      0 & 0 & e^{+i\theta/2} & 0 \\
      0 & 0 & 0 & e^{-i\theta/2}
   \end{pmatrix} \>,
\end{split}
\end{equation}
so that $\hat{\psi}(x) = S(\theta,\eta) \, \hat{\phi}(x)$ and
from Eq.~\eqref{h.e:psiIII}, we find
\begin{equation}\label{h.e:phiexpansion}
   \hat{\phi}(x)
   =
   S_{\rho}^{-1}(\theta) \, \hat{\phi}'(x)
   =
   \sum_{\bk} \sum_{\lambda = \pm 1}
   \hat{A}_{\bk}^{(\lambda)} \,
   \phi_{\bk}^{(\lambda)}(x) \>,
\end{equation}
where
\begin{equation}\label{h.e:phimodedef}
   \phi_{\bk}^{(\lambda)}(x)
   \defby
   e^{i \keta \eta} \,
   \begin{pmatrix}
      \phi_{(+);k}^{(\lambda)}(\tau) \,
      \chi_{k_m,+h}^{\phantom{(}}(\rho,\theta)
      \\
      \phi_{(-);k}^{(\lambda)}(\tau) \,
      \chi_{k_m,-h}^{\phantom{(}}(\rho,\theta)
   \end{pmatrix} \>,
\end{equation}
and where
\begin{equation}\label{h.e:chirhodef}
   \chi_{k_m,h}^{\phantom{(}}(\rho,\theta)
   =
   \frac{e^{i ( m + 1/2 ) \, \theta }}{\sqrt{2}}
   \begin{pmatrix}
      J_{m}(\kperp \rho)
      \\
      - h \, J_{m+1}(\kperp \rho)
   \end{pmatrix} \>.
\end{equation}
in agreement with the field expansion given in
Eqs.~\eqref{e:phiexpansion}, \eqref{e:phimodedef}, and
\eqref{e:chirhodef} in Sec.~\ref{ss:modeexpansion}.  We have
shown here that the separation of variables method for the
Dirac equation we used in Sec.~\ref{ss:modeexpansion} can
easily be understood as an expansion of transverse helicity
eigenvectors in boost-invariant coordinates.

The transverse helicity eigenvectors given in
Eq.~\eqref{h.e:chirhodef} satisfy the eigenvalue equation,
\begin{multline}\label{h.e:detranshel}
   \biggl [ \,
   i \, \sigma_y \,
   \Bigl ( \,
      \partial_{\rho} + \frac{1}{2 \rho} \,
   \Bigr )
   +
   i \, \sigma_x \, \frac{\partial_{\theta}}{\rho} \,
   \biggr ] \,
   \chi_{k_m,h}(\rho,\theta)
   \\
   =
   h \, \kperp \,
   \chi_{k_m,h}(\rho,\theta) \>,
\end{multline}
are normalized,
\begin{multline}\label{h.e:chinorm}
   \int_{0}^{\infty} \rho \rd \rho
   \int_{0}^{2\pi} \rd \theta \>
   \chi_{k_m,h}^{\phantom\dagger}(\rho,\theta) \,
   \chi_{k'_m,h'}^{\phantom\dagger}(\rho,\theta)
   \\
   =
   \delta_{h,h'} \,
   \delta_{m,m'} \,
   \frac{\delta(\kperp - \kperp')}{\sqrt{\kperp \kperp'}} \>,
\end{multline}
and complete
\begin{multline}\label{h.e:chicomp}
   \int_{0}^{\infty} \frac{\kperp \rd \kperp}{2\pi}
   \sum_{m=-\infty}^{+\infty}
   \sum_{h = \pm 1}
   \chi_{k_m,h}^{\phantom\dagger}(\rho,\theta) \,
   \chi_{k_m,h}^{\dagger}(\rho',\theta')
   \\
   =
   \delta(\theta - \theta') \,
   \frac{\delta(\rho - \rho')}{\sqrt{\rho \rho'}} \>.
\end{multline}

%
%
\section{Adiabatic expansion of solutions of the Dirac equation}
\label{s:adiabatic}

In this section, we find an adiabatic expansion of the positive
energy solutions of the Dirac equation for a slowly varying
field $A(\tau)$.  It is simplest to obtain an adiabatic
expansion of the polarization vector
$\smash{\bP_k^{(\lambda)}(\tau)}$, which we introduced in
Sec.~\ref{ss:modeexpansion}.  The equation of motion of the
polarization vector was given in Eq.~\eqref{e:eompol} as
\begin{equation}\label{pv.e:polde}
   \dot{\bP}_k^{(\lambda)}(\tau)
   =
   2 \,
   \bk_{k}(\tau)
   \times
   \bP_k^{(\lambda)}(\tau) \>,
\end{equation}
where $\bk_k(\tau)$ is given by
\begin{equation}\label{pv.e:bkdef}
   \bk_k(\tau)
   =
   \peta(\tau) \, \be_1
   +
   h \kperp \, \be_2
   +
   M \, \be_3 \>.
\end{equation}
The initial condition at $\tau=\tau_0$ is given in Eq.~\eqref{e:bp0def} as
\begin{equation}\label{pv.e:bp0def}
   \bP_k^{(\lambda)}(\tau_0)
   \defby
   \bP_{0;k}^{(\lambda)}
   =
   \lambda \, \bk_{0;k} / \omega_{0;k} \>.
\end{equation}
For slowly varying values of $\peta(\tau)$,
$\bP_k^{(\lambda)}(\tau)$ simply precesses about the slowly
varying value of $\bk_k(\tau)$.  In order to count derivatives
with respect to $\tau$, let us put
\begin{equation}\label{pv.e:epsilondef}
   \partial_{\tau}
   \mapsto
   \epsilon \, \partial_{\tau} \>.
\end{equation}
We next expand $\bP_k^{(\lambda)}(\tau)$ in powers of $\epsilon$ by writing
\begin{equation}\label{pv.e:pexpand}
\begin{split}
   \bP
   &=
   \bP^{(0)}
   +
   \epsilon \, \bP^{(1)}
   +
   \epsilon^2 \, \bP^{(2)}
   +
   \dotsb \>,
   \\
   \dot{\bP}
   &=
   \epsilon \, \dot{\bP}^{(0)}
   +
   \epsilon^2 \, \dot{\bP}^{(1)}
   +
   \epsilon^3 \, \dot{\bP}^{(2)}
   +
   \dotsb \>,
\end{split}
\end{equation}
Here and in the following, we omit momentum and time
dependencies and the $(\lambda)$ label.  The superscript now
counts powers of $\epsilon$ and the dot refers to derivatives
with respect to $\tau$.  So substitution of
\eqref{pv.e:pexpand} into Eq.~\eqref{pv.e:polde} becomes
\begin{multline}\label{pv.e:polexpde}
   \epsilon \, \dot{\bP}^{(0)}
   +
   \epsilon^2 \, \dot{\bP}^{(1)}
   +
   \epsilon^3 \, \dot{\bP}^{(2)}
   +
   \dotsb
   \\
   =
   2 \,\bk
   \times
   \bigl [ \,
      \bP^{(0)}
      +
      \epsilon \, \bP^{(1)}
      +
      \epsilon^2 \, \bP^{(2)}
      +
      \dotsb \,
   \bigr ] \>.
\end{multline}
Equating equal powers of $\epsilon$ gives
\begin{subequations}\label{pv.e:allpterms}
\begin{align}
   2 \, \bk \times \bP^{(0)}
   &=
   0 \>,
   \label{pv.e:pta} \\
   2 \, \bk \times \bP^{(1)}
   &=
   \dot{\bP}^{(0)} \>,
   \label{pv.e:ptb} \\
   2 \, \bk \times \bP^{(2)}
   &=
   \dot{\bP}^{(1)} \>,
   \label{pv.e:ptc} \\
   2 \, \bk \times \bP^{(3)}
   &=
   \dot{\bP}^{(2)} \>, \qquad\text{etc $\dotsb$}
   \label{pv.e:ptd}
\end{align}
\end{subequations}
Let us introduce transverse and longitudinal components of the
polarization vector by writing $\bP^{(\kappa)} =
\bP^{(\kappa)}_{T} + \bP^{(\kappa)}_{L}$, where $\bk \cdot
\bP^{(\kappa)}_{T} = 0$ and $\bk \times \bP^{(\kappa)}_{L} =
0$.  So Eqs.~\eqref{pv.e:allpterms} determine only the
transverse components of the polarization vector.  The
longitudinal portion is then fixed by the normalization
requirement, as we will see below.

From Eq.~\eqref{pv.e:pta}, $\bP^{(0)}$ is entirely longitudinal
and  has the normalized solution
\begin{equation}\label{pv.e:p0sol}
   \bP^{(0)}
   =
   \bk / \omega \>,
   \quad
   \omega
   =
   | \bk |
   =
   \sqrt{ \peta^2 + \kperp^2 + M^2 } \>.
\end{equation}
So $\bP^{(0)} = \bP_0$.  So we have
\begin{equation}\label{pv.e:dotp0}
   \dot{\bP}^{(0)}
   =
   \frac{\dot{\bk}}{\omega}
   -
   \frac{\bk \, \dot{\omega} }{\omega^2}
   =
   \frac{\dot{\bk}}{\omega}
   -
   \frac{ \bk \, ( \dot{\bk} \cdot \bk ) }{ \omega^3 }
   =
   \frac{ \bk \times ( \dot{\bk} \times \bk ) }{ \omega^3 }  \>.
\end{equation}
Then Eq.~\eqref{pv.e:ptb} becomes
\begin{equation}\label{pv.e:p1de}
   2 \, \bk \times \bP^{(1)}
   =
   \frac{ \bk \times ( \dot{\bk} \times \bk ) }{ \omega^3 } \>,
\end{equation}
so the transverse component of $\bP^{(1)}$ is given by
\begin{equation}\label{pv.e:p1T}
   \bP^{(1)}_{T}
   =
   \frac{ \dot{\bk} \times \bk }{ 2 \, \omega^3 } \>.
\end{equation}
We will choose the longitudinal component $\bP^{(1)}_{L} = 0$, so that
\begin{equation}\label{pv.e:p1sol}
   \bP^{(1)}
   =
   \frac{ \dot{\bk} \times \bk }{ 2 \, \omega^3 }
   =
   -
   \frac{\dpeta}{ 2 \, \omega^3 } \,
   \bigl ( \,
      M \, \be_2
      -
      h \kperp \, \be_3 \,
   \bigr ) \>.
\end{equation}
Then to first order, $\bP = \bP^{(0)} + \epsilon \, \bP^{(1)}$,
the polarization vector is normalized to this order, since
\begin{equation}\label{pv.e:normp1}
   P^2
   =
   P_0^2
   +
   2 \epsilon \, \bP^{(0)} \cdot \bP^{(1)}
   =
   1 \>.
\end{equation}
From \eqref{pv.e:p1sol}, we find
\begin{equation}\label{pv.e:dotp1}
\begin{split}
   \dot{\bP}^{(1)}
   &=
   \frac{ \ddot{\bk} \times \bk }{ 2 \, \omega^3 }
   -
   \frac{ 3 \, ( \dot{\bk} \times \bk ) \, \dot{\omega} }
        { 2 \, \omega^4 }
   \\
   &=
   \frac{ \ddot{\bk} \times \bk }{ 2 \, \omega^3 }
   -
   \frac{ 3 \, ( \dot{\bk} \times \bk ) \,
          ( \dot{\bk} \cdot \bk ) }
        { 2 \, \omega^5 }
   \\
   &=
   \bk \times
   \Bigl [ \,
   \frac{ 3 \, ( \dot{\bk} \cdot \bk ) \,  \dot{\bk}
          - \omega^2 \, \ddot{\bk} }
        { 2 \, \omega^5 } \,
   \Bigr ] \>.
\end{split}
\end{equation}
Then Eq.~\eqref{pv.e:ptc} becomes
\begin{equation}\label{pv.e:p2de}
   2 \, \bk \times \bP^{(2)}
   =
   \bk \times
   \Bigl [ \,
   \frac{ 3 \, ( \dot{\bk} \cdot \bk ) \,  \dot{\bk}
          - \omega^2 \, \ddot{\bk} }
        { 2 \, \omega^5 } \,
   \Bigr ] \>,
\end{equation}
so adding a longitudinal part to $\bP^{(2)}$, we find
\begin{equation}\label{pv.e:p2sol}
   \bP^{(2)}
   =
   \frac{ 3 \, ( \dot{\bk} \cdot \bk ) \,  \dot{\bk}
          - \omega^2 \, \ddot{\bk} }
        { 4 \, \omega^5 }
   +
   \mathcal{N}_2 \, \bk \>,
\end{equation}
where $\mathcal{N}_2$ is to be fixed by the normalization
requirement.  From the expansion \eqref{pv.e:polexpde}, we find
to second order
\begin{equation}\label{pv.e:pnorm}
   P^2
   =
   1
   +
   \epsilon^2 \,
   \bigl [ \,
      P^{(1)\,2}
      +
      2 \, \bP^{(0)} \cdot \bP^{(2)} \,
   \bigr ]
   +
   \dotsb
   =
   1 \>.
\end{equation}
So we want to choose $\mathcal{N}_2$ such that $P^{(1)\,2} + 2
\, \bP^{(0)} \cdot \bP^{(2)} = 0$.  This gives the equation
\begin{equation}\label{pv.e:Nchoice}
\begin{split}
   \bP^{(0)} \cdot \bP^{(2)}
   &=
   \frac{ 3 \, ( \dot{\bk} \cdot \bk )^2
          - \omega^2 \, \ddot{\bk} \cdot \bk }
        { 4 \, \omega^6 }
   +
   \mathcal{N}_2 \, \omega
   \\
   &=
   -
   \frac{1}{2} \, P^{(1)\,2}
   =
   -
   \frac{ | \dot{\bk} \times \bk |^2 }{ 8 \, \omega^6 } \>,
\end{split}
\end{equation}
from which we find
\begin{equation}\label{pv.e:Nvalue}
\begin{split}
   \mathcal{N}_2
   &=
   -
   \frac{ 3 \, ( \dot{\bk} \cdot \bk )^2
          - \omega^2 \, \ddot{\bk} \cdot \bk }
        { 4 \, \omega^7 }
   -
   \frac{ | \dot{\bk} |^2 \, \omega^2 - ( \bk \cdot \dot{\bk} )^2 }
        { 8 \, \omega^7 }
   \\
   &=
   -
   \frac{1}{8} \, \frac{\dpeta^2}{\omega^5}
   +
   \frac{1}{4} \, \frac{\peta \, \ddpeta}{\omega^5}
   -
   \frac{5}{8} \, \frac{\peta^2 \, \dpeta^2}{\omega^7} \>.
\end{split}
\end{equation}
\bigskip
So to second adiabatic order, the polarization vector is given by
\begin{align}
   \bP
   &=
   \bP^{(0)}
   +
   \epsilon \, \bP^{(1)}
   +
   \epsilon^2 \, \bP^{(2)}
   +
   \dotsb
   \label{pv.e:psecondorder} \\
   &=
   \frac{\bk}{\omega}
   +
   \epsilon \,
   \frac{ \dot{\bk} \times \bk }{ 2 \, \omega^3 }
   +
   \epsilon^2 \,
   \Bigl [ \,
   \frac{ 3 \, ( \dot{\bk} \cdot \bk ) \,  \dot{\bk}
          - \omega^2 \, \ddot{\bk} }
        { 4 \, \omega^5 }
   +
   \mathcal{N}_2 \, \bk \,
   \Bigr ]
   +
   \dotsb
   \notag
\end{align}
In component form, we find
\begin{subequations}\label{pv.e:pvector}
\begin{align}
   P_1
   &=
   \frac{\peta}{\omega}
   -
   \epsilon^2 \, ( \, \kperp^2 + M^2 \, ) \,
   \Bigl ( \,
      \frac{1}{4} \,
      \frac{ \ddpeta }{ \omega^5 }
      -
      \frac{5}{8} \,
      \frac{ \peta \, \dpeta^2 }{ \omega^7 } \,
   \Bigr )
   +
   \dotsb \>,
   \label{ad.e:P1comp}
   \\
   P_2
   &=
   \frac{h \kperp}{\omega}
   -
   \epsilon \, M \,
   \frac{\dpeta}{2 \, \omega^3}
   \label{ad.e:P2comp} \\ &\quad
   +
   \epsilon^2 \, h \kperp \,
   \Bigl ( \,
   -
   \frac{1}{8} \,
   \frac{\dpeta^2}{\omega^5}
   +
   \frac{1}{4} \,
   \frac{\peta \, \ddpeta}{\omega^5}
   -
   \frac{5}{8} \,
   \frac{\peta^2 \, \dpeta^2}{\omega^7} \,
   \Bigr )
   +
   \dotsb \>,
   \notag \\
   P_3
   &=
   \frac{M}{\omega}
   +
   \epsilon \, h \kperp \,
   \frac{\dpeta}{2 \, \omega^3}
   \label{ad.e:P3comp} \\ &\quad
   +
   \epsilon^2 \, M \,
   \Bigl ( \,
   -
   \frac{1}{8} \,
   \frac{\dpeta^2}{\omega^5}
   +
   \frac{1}{4} \,
   \frac{\peta \, \ddpeta}{\omega^5}
   -
   \frac{5}{8} \,
   \frac{\peta^2 \, \dpeta^2}{\omega^7} \,
   \Bigr )
   +
   \dotsb \>,
   \notag
\end{align}
\end{subequations}
which completes the adiabatic analysis used in this paper.

\vfill

%
%
\bibliography{johns}
%
%

%
%
\end{document}